\documentclass[10pt]{iopart}
\usepackage{amssymb}

\usepackage{iopams}
\usepackage{setstack}
\usepackage[square,numbers,sort & compress,comma]{natbib}
\usepackage{graphicx}
\usepackage{appendix}
\usepackage{multicol}
\usepackage{graphicx}
\usepackage{subfigure}
\usepackage{url}
\usepackage{caption}

\newtheorem{theorem}{Theorem}
\input{tcilatex}
\def\newblock{\hskip .11em plus .33em minus .07em}
\newcommand{\stackunder} [1] {\underset{#1}}
\begin{document}

\title{Mixing, entropy and competition }
\author{A.Y. Klimenko}
\address{The University of Queensland, SoMME, QLD 4072, Australia \\ Email: klimenko@mech.uq.edu.au}

\begin{abstract}
Non-traditional thermodynamics, applied to
random behaviour associated with turbulence, mixing and competition, is
reviewed and analysed. Competitive mixing represents a general framework for
the study of generic properties of competitive systems and can be used to
model a wide class of non-equilibrium phenomena ranging from turbulent
premixed flames and invasion waves to complex competitive systems. We
demonstrate consistency of the general principles of competition with
thermodynamic description, review and analyse the related entropy concepts
and introduce the corresponding competitive H-theorem. A competitive system
can be characterised by a thermodynamic quantity --- competitive potential
--- which determines the likely direction of evolution of the system.
Contested resources tend to move between systems from lower to higher values
of the competitive potential. There is, however, an important difference
between conventional thermodynamics and competitive thermodynamics. While
conventional thermodynamics is constrained by its zeroth law and is
fundamentally transitive, the transitivity of competitive thermodynamics
depends on the transitivity of the competition rules. Intransitivities are
common in the real world and are responsible for complex behaviour in
competitive systems.

This work follows the ideas and methods that are originated in analysis of turbulent
combustion but reviews a much broader scope of issues linked to mixing and
competition, including thermodynamic characterisation of complex competitive
systems with self-organisation. The approach presented here is interdisciplinary and is
addressed to a general educated reader, while the mathematical details can
be found in the Appendices.

\end{abstract}

\bigskip
\bigskip
\bigskip

\bigskip
\bigskip
\bigskip

\bigskip
\bigskip
\bigskip

\bigskip
\bigskip
\bigskip

-----------------------------------

{\small Published: Phys. Scr. 85 (2012) 068201 (29pp)}


\maketitle



\section{Introduction}

Thermodynamics allows for a concise description of complex stochastic
systems, determining an overall trend behind a large number of random events
and offering insightful generalisations. The success of classical
thermodynamics is largely based on recognising and postulating
irreversibility of the surrounding world that on one hand represents an
obvious fact and on the other hand still awaits explanation from the first
principles of physics. The second law of thermodynamics, which predicts
irreversible increase of entropy --- the key thermodynamic quantity serving
as a measure of chaotic uncertainty --- is equally applicable to a small
combustor and to stars and galaxies.

This remarkable success of thermodynamics can not hide from us its major
difficulty - our world appears to be much more complicated and much less
chaotic than generally might be inferred from the second law. It is well
known that complex non-equilibrium stochastic processes tend to display a
significant level of regularity along with randomness \cite{Prig1977}. In
non-equilibrium phenomena, the production of physical entropy is typically
high, in perfect agreement with the laws of thermodynamic. Although no
direct violation of the laws of thermodynamics is known, thermodynamics
struggles to explain complexity, which is often observed in essentially
non-equilibrium phenomena: turbulent mixing and combustion as well as
evolution of life forms may serve as typical examples. The entropy of
turbulent fluctuations does not seem to be maximal and the same applies to
entropies characterising distributions in other complex non-equilibrium
processes. These entropies have similarities with but are not the same as
the molecular entropy, which characterises disorder of molecular movements
and is subject to the laws of thermodynamics. We use the term \textit{%
apparent entropy} to distinguish entropy-like quantities from the molecular
entropy.

The present work reviews the use of entropy in the analysis of turbulence,
turbulent mixing and combustion and shows that the term ``entropy'' is
commonly used to denote both apparent entropy and molecular entropy. The
same trend can be observed across other disciplines. In principle, the use
of apparent entropy may or may not imply the existence of underlying
thermodynamics. The existence of apparent thermodynamics associated with
mixing is of prime interest for this work. Thermodynamic description is a
very general methodology involving abstract theories, i.e. theories not
directly linked to the dynamics of molecules. The general theory of \textit{%
Gibbs measures} \cite{GibbsMeasure80} and the axiomatic thermodynamic theory 
\cite{EntOrd2003}, introducing entropy on the basis of ordering of
thermodynamic states by Caratheodory's \textit{adiabatic accessibility} \cite
{Cara1909}, should be mentioned in this respect.

Competitive systems, which are typically associated with complex stochastic
behaviour, are common in the real world. \textit{Abstract competition},
which studies generic principles of competition in their most abstract form,
can be interpreted as a form of mixing \cite{K-PS2010}. This mixing, which
is called \textit{competitive mixing}, can be used to characterise various
processes: turbulent combustion, invasion waves and other related phenomena 
\cite{KP2012}. Unlike conventional conservative mixing, competitive mixing
can display complex behaviour with sophisticated interdependencies. After
reviewing existing publications and taking into account a number of theorems
presented in the Appendices, we demonstrate that competitive systems do
allow for a thermodynamic description. The implications of this
demonstration are profound: the evolution of competitive systems occurs in a
stochastic manner but in agreement with \textit{competitive thermodynamics}.
A competitive system can be characterised by competitive entropy and by a
new thermodynamic quantity --- the \textit{competitive potential} --- which
determines the likely direction of evolution of the system and is analogous
to chemical potential in conventional thermodynamics (taken with the
opposite sign). Following in the footsteps of classical thermodynamics,
competitive thermodynamics recognises the obvious trend towards more
competitive states while the details of the mechanism behind the competition
rules may remain unknown.

Competitive thermodynamics, while answering many existing questions, poses
several new ones. A conventional thermodynamic system evolves towards
equilibrium and, once the global equilibrium is reached remains in this
state indefinitely. Competitive systems tend to display much more complex
and unending pattern of evolution --- how this can be consistent with a
thermodynamic description? The answer lies in the details. Competitive
thermodynamics has a principal difference with conventional thermodynamics:
transitivity of competitive thermodynamics can not be taken for granted.
While conventional thermodynamics is constrained by its zeroth law and is
fundamentally transitive, the transitivity of competitive thermodynamics
depends on the transitivity of the competition rules. \textit{%
Intransitivities} are not only possible in real competitive systems, but
seem to be quite common. Unlike chemical potential or temperature, which can
be assigned absolute values, the competitive potential becomes relative and
this removes the rigid constraints of conventional thermodynamics and
introduces complex patens into evolution. Intransitivity, which has long
been known in science under the name of the \textit{Condorcet paradox} \cite
{Cond1785} and has traditionally been considered as something paradoxical,
abnormal or unwanted, is viewed here as a common property of nature.

The approach presented here is derived from the long-standing tradition of
modelling turbulent reacting flows repeatedly reviewed in publications \cite
{Pope85,Williams85,Kuznetsov-S-90,Maas1996,KB99,Pope-00,Fox-2003,Heinz-2003,Pitsch2006,Haworth2009,book2011}%
. The rapid development over the last few decades of computational models
designed for the simulation of transport, reaction and dispersion in
turbulent flows has led to a wide use of \textit{Pope particles }\cite
{Pope85,KC-PopeFTC}\textit{.} These notional particles move in physical
space and posses a set of properties that can be changed due to 1) kinetic
evolution within each particle and 2) mixing exchanges between the
particles. With the introduction of competitive mixing, Pope particles can
be viewed not only as being effective tools for modelling of turbulent
reacting flows but also as universal building blocks for a wide class of
models that can simulate complex behaviour. Conventional conservative mixing
does not result in significant stochastic interdependencies between the
particles and a system of many particles can be characterised by a single
one-particle pdf. If conventional mixing is replaced by competitive mixing,
these interdependencies may become significant, dramatically increasing the
effective dimensionality and complexity of the simulations. Competitive
mixing naturally appears in simulations of turbulent premixed flames \cite
{KP2012}, which are driven by the forces of conventional thermodynamics. We
review these applications and take the logical step of extending these
thermodynamic descriptions to more complex competitive systems. There is a
large number of publications dedicated to different aspects of complexity,
for example, algorithmic (Kolmogorov-Chaitin)\ complexity and algorithmic
entropy \cite{AlgEntrop2009}, \textit{complex adaptive systems (CAS)} \cite
{G-M1994,CAS2006} and evolution of complexity \cite{CompEvol1999}.

The phenomena we consider display a combination of chaotic and ordered
behaviour. Entropy, which is conventionally used to characterise the balance
of order and disorder, has been repeatedly applied to the analysis of
non-equilibrium systems in general \cite
{PrigGlan,MAP1979,Ziegler1983,Ftheor2002,MEP2006,SWENSON2009} and turbulent
flows in particular \cite{Pope-79c,Dupree1992,SANCHO1994,
Falk2004,Celani2005,Attard2007,Duplat2008}. Our treatment of non-equilibrium
processes is based on introducing non-conventional or apparent
thermodynamics, which is not analogous to but still may have some links with
the principles of non-equilibrium thermodynamics (i.e. entropy production
principles of Prigogine \cite{PrigGlan} and Ziegler \cite{Ziegler1983} and
the \textit{fluctuation theorem} \cite{Ftheor2002}). This work follows the\
application of the concept of entropy to semi-autonomous elements, which in
most cases can be effectively represented by Pope particles.

In accordance with the ``Turbulent Mixing and Beyond'' tradition, the review
starts from the methods used in modelling of turbulent mixing and combustion
and then extends these methods beyond turbulence to mixing and competitive
systems of a general nature. The possibility of thermodynamic description is
sought and, in many cases, found for these systems. While the thermodynamic
description of mixing, both conservative and competitive, is the focus of
the present work, the other areas are covered as necessary but only to the
extent of their relevance to the main topic.

The review is divided into 8 sections and 4 appendices. The sections present
the following material:

\begin{itemize}
\item  \textbf{Section 2} defines entropy for systems of notional particles.
The defined entropy involves two major components: configurational (related
to collective particle disorder) and potential (related to the state of each
particle). The use of particle entropy in context of turbulent flows is
subsequently reviewed.

\item  \textbf{Section 3} analyses the effect of conservative and
competitive mixing on entropy. Conventional conservative mixing models,
which are commonly used in modelling of transport and reactions in turbulent
flows, are reviewed and the condition that enforces conservative mixing to
be entropy-consistent is presented. The conventional entropy of premixed
combustion is extended (as apparent entropy) to become a common property of
systems involving competitive mixing.

\item  \textbf{Section 4} explores the fundamental link between ordering,
ranking and entropy. The analysis is directed at competitive mixing but the
related methodologies developed in other disciplines (adiabatic
accessibility and economic utility) are also reviewed.

\item  \textbf{Section 5} analyses the behavior of systems with transitive
competition and shows that it is thermodynamically consistent. The
thermodynamic analogy is especially strong for the class of mutations that
is called Gibbs mutations by analogy with Gibbs measures. A transitive
competitive system tend to promptly reach a quasi-equilibrium state and then
slowly drift in the direction of increasing competitiveness. Both processes
are characterised by increase in apparent entropy.

\item  \textbf{Section 6} investigates a more complex case of intransitive
competition. Thermodynamic considerations can be applicable if intransitive
competition retains some transitive properties. The applicability of
competitive thermodynamics to general intransitive systems, which may
involve competitive cooperation and other forms of complex behaviour, is
also explored.

\item  \textbf{Section 7} gives several examples of intransitive behaviour
including intransitivity in turbulent flows, in chemical reactions and in a
generic competitive system displaying complex behaviour.

\item  \textbf{Section 8} outlines the main conclusions for this work.

\item  \textbf{Apendices} present useful mathematical details and additional
material, which is not available in the published literature but is
essential to this review:

\begin{itemize}
\item  \textbf{Appendix A} gives a brief summary of the related mathematical
results.

\item  \textbf{Appendix B} generalises rankings for preferential mixing

\item  \textbf{Appendix C} introduces Gibbs and near-Gibbs mutations and
explores their relations with Markov processes, Gibbs measures and the
fluctuation theorem.

\item  \textbf{Appendix D} presents governing equations, general theorems
and analysis of some special cases for evolution of competitive systems.
\end{itemize}
\end{itemize}

\section{Entropy of particle systems}

Although entropy was introduced in thermodynamics and statistical physics by
Clausius, Boltzmann and Gibbs as a specific, heat-related property of large
systems of molecules, the modern use of this term ranges from rigorous
extensions of the concept of entropy (such as Shannon's entropy in
information theory) to relatively vague and intuitive interpretations (such
as social entropy). The more general view of entropy, which takes its origin
in Shannon's famous work \cite{Shan1951}, sees entropy as a property
characterising disorder of stochastic behaviour in general. In the present
we understand entropy as a quantity which displays (or is expected to
display) behavior similar to that of the molecular entropy. This entropy,
though, does not necessarily coincide with the molecular entropy used in
conventional thermodynamics and the word \textit{apparent} is used whenever
it is necessary to stress this difference. Thermodynamic quantities
introduced for competitive systems can be also named as \textit{competitive}%
. In this section, common definitions of entropy for a system of notional
particles are considered.

\subsection{Configurational entropy and potential entropy}

Consider $n$ notional stochastic particles, where each of these particle is
characterised by a vector $\mathbf{X}=(X^{(1)},...,X^{(k_{d})}).$\ The joint
probability distribution of these particles is denoted by $P_{n}=P_{n}(%
\mathbf{X}_{1},...,\mathbf{X}_{n})$. The Gibbs entropy is introduced as a
statistical sum (or integral) over all possible states of this system 
\begin{equation}
\bar{S}=\int_{\infty }\left( -P_{n}\ln \left( \frac{P_{n}}{A_{n}}\right)
+P_{n}s_{n}\right) d\mathbf{X}_{1}...d\mathbf{X}_{n},
\end{equation}
where $A_{n}$ can be interpreted as a priori statistical weight
characterising effective volumes in the parameter space. This definition is
conventional \cite{Falk2004} but includes an additional term $P_{n}s_{n},$
which is considered below. If the particles are (or can be treated as)
statistically independent then the joint pdf (probability density function)
is decomposed into single-particle pdfs 
\begin{equation}
P_{n}=P_{n}(\mathbf{X}_{1},...,\mathbf{X}_{n})=P(\mathbf{X}_{1})...P(\mathbf{%
X}_{n}),
\end{equation}
and the equation for entropy takes the form of Boltzmann entropy 
\begin{equation}
\bar{S}=\;\stackunder{\bar{S}_{c}}{\underbrace{-n\int_{\infty }P(\mathbf{X}%
)\ln \left( \frac{P(\mathbf{X})}{A(\mathbf{X})}\right) d\mathbf{X}}}\;+\;%
\stackunder{\bar{S}_{f}}{\underbrace{n\int_{\infty }P(\mathbf{X})s(\mathbf{X}%
)d\mathbf{X}}}  \label{S-Bol}
\end{equation}
The first term $\bar{S}_{c},$ which is called here \textit{configurational
entropy}, is related to the stochastic nature of the particles distribution
while the second term $\bar{S}_{f}$, which is called \textit{potential} 
\textit{entropy,} is related to the particle state and characterised by the
entropy potential $s(\mathbf{X})$. If we interpret the particles as
computational objects, the configurational entropy is the same as Shannon
entropy of variable $\mathbf{X}$. It is arguable that, if the particles are
not distinguishable, the value $\ln (n!)\approx n\ln (n)$ needs to be
deducted from (\ref{S-Bol}). This however does not affect our considerations
since the number of particles is kept constant. The entropy $\bar{S}$ can be
interpreted as free entropy, defined as $\bar{S}=-G/T$ where $G$ is the
Gibbs (or Helmholtz) free energy and $T$ is the absolute temperature
measured in energy units. In this case the physical interpretation of $s(%
\mathbf{X})$ is most transparent and linked to free energy of each state $%
\mathbf{X}$. The distinction between configurational and non-configurational
free energies is commonly used in thermodynamic modelling \cite{Saulov2009a}%
. In the present work, we do not restrict our consideration to a specific
interpretation of $\bar{S}$. Inclusion of a priori statistical weight $A(%
\mathbf{X})$ makes the definition of entropy invariant with respect to
replacements of variables $\mathbf{X}$.

The \textit{Gibbs entropy} and \textit{Boltzmann entropy} are equivalent
only if the particles are independent. The particles may display some
dependence in case of conventional conservative mixing \cite{Klim2009-1} but
these dependencies are typically small. As discussed in the following
sections, complex particle behavior, which can be observed in the case of
competitive mixing, may be accompanied by significant particle dependencies
and substantial differences between the two definitions. In this case,
however, Gibbs entropy becomes computationally intractable since the sum is
to be evaluated over all alternative realizations in the overall composition
space of very large dimension $n\times k_{d}.$ Typically, these alternative
realizations remain unknown in computations while the whole ensemble of
realisations may be difficult to define for complex systems. Our analysis is
largely based on Boltzmann entropy, which is evaluated using the discrete
representation of the single-particle pdf $P(\mathbf{X})$ by the current
distribution of $n$ particles, where $n$ is assumed to be large. Note that
in complex systems the current distribution may fluctuate even if $n$ is
large.

\subsection{Entropy of Pope particles}

We now consider Pope particles and distinguish location of the particle
denoted by $\mathbf{x}=(x_{1},...,x_{k_{x}})$ and particle properties
denoted by $\mathbf{y}=(y_{1},...,y_{k_{y}}),$ that is $\mathbf{X}=(\mathbf{x%
},\mathbf{y})$ and $k_{d}=k_{x}+k_{y}$. The variables $x_{1},...,x_{k_{x}}$
represent physical coordinates and, possibly, other reference variables such
as those used in MMC mixing \cite{KP03,book2011CK}. The coordinates $\mathbf{%
x}$ are conventionally governed by a Markov diffusion process while the
particle properties $\mathbf{y}$ change due to mixing and, possibly,
chemical reactions. With this distinction drawn between the physical
coordinates $\mathbf{x}$ and particle properties $\mathbf{y,}$ we assume
that entropy of the particle state $s$ is dependent on particle properties $%
\mathbf{y}$ but not on particle coordinates $\mathbf{x}$ (that is particles
are not placed in any force field acting in physical space). The entropy can
be divided into volumetric and local $S=S(\mathbf{x})$ components according
to

\[
\bar{S}=n\int_{\infty }\left( S(\mathbf{x})-\ln (p(\mathbf{x}))\right) p(%
\mathbf{x})d\mathbf{x}
\]
\[
S(\mathbf{x})=\int_{\infty }p(\mathbf{y|x})s(\mathbf{y})d\mathbf{y-}%
\int_{\infty }p(\mathbf{y|x})\ln \left( \frac{p(\mathbf{y|x})}{A(\mathbf{y,x}%
)}\right) d\mathbf{y}\;
\]
\bigskip The one-particle pdf $P=P\left( \mathbf{y},\mathbf{x},t\right) $
governing distribution of Pope particles\ satisfies the equation \cite
{Pope85, Klim2009-1} 
\begin{equation}
\frac{\partial P}{\partial t}+\func{div}\left( \mathbf{v}P\right) -D\nabla
^{2}P+\sum_{j}\frac{\partial W^{(j)}P}{\partial y^{(j)}}=\left[ \frac{dP}{dt}%
\right] _{\limfunc{mix}},  \label{pdfx}
\end{equation}
where $\mathbf{v}$ is the velocity in physical space $\mathbf{x}$, $D$ is
the diffusion coefficient in physical space, $W^{(j)}$ is the reaction rate
and the term on right hand side symbolically represents the effect of
mixing. After some conventional manipulations, differentiating equation (\ref
{S-Bol}) results in 
\[
\frac{d\bar{S}}{dt}=n\stackunder{\infty }{\int \int }\left( \digamma
_{x}+\sum_{j}\frac{\partial W^{(j)}}{\partial y^{(j)}}-\ln (P)\left[ \frac{dP%
}{dt}\right] _{\limfunc{mix}}\right) d\mathbf{x}d\mathbf{y+}
\]
\begin{equation}
+n\stackunder{\infty }{\int \int }\left( P\sum_{j}W^{(j)}\frac{\partial s(%
\mathbf{y})}{\partial y^{(j)}}+s(\mathbf{y})\left[ \frac{dP}{dt}\right] _{%
\limfunc{mix}}\right) d\mathbf{y}d\mathbf{x,}
\end{equation}
where 
\[
\digamma _{x}=P\func{div}\left( \mathbf{v}\right) +D\frac{\left( \nabla
P\right) ^{2}}{P}
\]
represents terms related to spatial inhomogeneity. The velocity divergence
term was previously derived and investigated by Falkovich and Fouxon \cite
{Falk2004}, who concluded that this term may result in entropy extraction
from the system to the environment. The effect of the second term is well
known -- this term contributes to generation of entropy \cite{MixControl2010}%
. In the present work we focus on the mixing term and its influence on
entropy and mainly restrict our attention to a spatially homogenous and
non-reacting case. The mixing operator is typically presumed to be \textit{%
localised} in $\mathbf{x}$-space and can be assumed to be\ \textit{%
non-preferential} with respect to $\mathbf{y}$. That is all particles are
selected for mixing from a given location $\mathbf{x}$ with equal
probability irrespective of their properties. The simplest mixing models are
non-preferential but, in principle, modelling can be improved by exercising
proper preferences during mixing.

\subsection{Maximal entropy distribution and competitive potential}

In the rest of the paper we denote $P(\mathbf{y|x})=f(\mathbf{y})$ and
consider only local characteristics so that the equation for the local
entropy takes the form 
\begin{equation}
S=\stackunder{S_{c}}{\;\underbrace{-\int_{\infty }f(\mathbf{y})\ln \left( 
\frac{f(\mathbf{y})}{A(\mathbf{y})}\right) d\mathbf{y}}}\;+\;\stackunder{%
S_{f}}{\underbrace{\int_{\infty }f(\mathbf{y})s(\mathbf{y})d\mathbf{y}}}
\label{Scs}
\end{equation}
With the use of the equilibrium function $f_{0}(\mathbf{y})$ defined by 
\begin{equation}
f_{0}(\mathbf{y})=\frac{A(\mathbf{y})}{Z}\exp \left( s(\mathbf{y})\right) ,
\label{eq-f0}
\end{equation}
where 
\begin{equation}
Z=\int_{\infty }A(\mathbf{y})\exp \left( s(\mathbf{y})\right) d\mathbf{y}
\label{eq-Z}
\end{equation}
is the partition function, the entropy equation takes the form 
\begin{equation}
S([f])\mathbf{=}\int_{\infty }f(\mathbf{y})\left( \ln (Z)-\ln \left( \frac{f(%
\mathbf{y})}{f_{0}(\mathbf{y})}\right) \right) d\mathbf{y}  \label{Sf}
\end{equation}
The notation $S([f])$ is used to emphasise that $S$ is a functional of the
distribution $f(\mathbf{y})$. Equation (\ref{Sf}) is similar to the
Kullback-Leibler divergence \cite{K-Ldiv}, that is known to achieve global
entropy maximum 
\[
S([f_{0}])=\ln (Z) 
\]
by the distribution $f(\mathbf{y})=f_{0}(\mathbf{y})$.

For competitive systems, we also introduce \textit{competitive potential} $%
\chi $ defined by 
\begin{equation}
\chi (\mathbf{y})=\frac{\delta S}{\delta f(\mathbf{y})}=-\ln \left( \frac{f(%
\mathbf{y})}{f_{0}(\mathbf{y})}\right) +\chi _{0},  \label{mu}
\end{equation}
where 
\begin{equation}
\chi _{0}=\ln (Z)-1  \label{mu0}
\end{equation}
is the competitive potential of the equilibrium state. The constant of unity
can be omitted from these equations. The a priori statistical weight $A(%
\mathbf{y})$ can be subject to different physical interpretations in
competitive systems but it seems most logical to link $A(\mathbf{y})$ to the
probability distributions assuming that competition is switched off (i.e. to
the a priori probability). Particles with fixed $\mathbf{y}$ can be treated
as reactants with potential $\chi $ while $\chi _{0}$ represents the
potential of the system composed from different equilibrated reactants. Note
that $f(\mathbf{y})$ should be multiplied by $n,$ if the number of particles 
$n$ may change. The competitive potential can be seen as a thermodynamic
quantity which is similar to the chemical potential of reacting systems
although $\chi $ is defined with the opposite sign. The sign of $\chi $ is
selected to avoid a direct conflict with the common-sense interpretation of
the expression of ``having a high competitive potential''. This change in
sign does not affect any physical properties of the system and is purely a
notational matter. The similarity of competitive and chemical potentials is
linked to the fact that it is the number of particles that is presumed to be
preserved in interactions. Entropy combined with preservation of energy
introduces the temperature. The particle systems considered here do not have
temperature as long as there is no associated energy-like quantity that is
conserved in mixing interactions.

\subsection{Entropy in studies of turbulent flows.}

In this section, we review the use of entropy in studies of turbulent flows
revealing that different physical quantities or different conditions may in
fact be implied when invoking this term. The molecular entropy has been used
on numerous occasions to construct models of turbulent flows. The following
examples indicate the wide scope of possible applications but, of course,
are not intended as a comprehensive review of all possible applications.
Molecular entropy can be used to characterise the spectrum of convective
turbulence \cite{Lvov1991}, ensure consistency of models with the laws of
thermodynamics \cite{RenPope2004}, control mixing processes \cite
{MixControl2010}, or model turbulent combustion \cite{CombModEntrop}. In the
last work \cite{CombModEntrop}, entropy is used in stochastic simulations as
a convenient progress variable that allows for effective reduction of the
chemical composition space \cite{BykovMaas2010}.

Production of molecular entropy is the key factor in two general principles
applicable to non-equilibrium dynamics: Prigogine's theorem of minimal
entropy production \cite{PrigGlan} and Ziegler's maximal entropy production
(MEP) principle \cite{Ziegler1983}. Despite the apparent contradiction,
these principles do not interfere with each other and both of the principles
are consequences of Onsager's reciprocal relations. Prigogine's theorem is
formulated for the specific conditions of a system asymptotically converging
to steady (but not necessarily equilibrium) state where entropy production
reaches its minimum while MEP is related to determining thermodynamic flows
for given thermodynamic forces and at a fixed moment of time. According to a
number of authors \cite{MAP1979,SWENSON2009}, MEP can also be viewed as a
very general principle: if a nonlinear system has several routes of moving
towards its equilibrium state, nature seems to prefer the route with maximal
entropy production. For example, turbulent flow is a more likely state than
laminar flow and the former has higher dissipation and higher entropy
production. This MEP principle seems to be very plausible and general but it
still needs qualifications of conditions and justification \cite{MAP2005}.
Ozawa et al \cite{Ozawa2001} analysed several different\ types of turbulent
flows and concluded that MEP is applicable to these flows. The probabilities
of positive and negative values of the entropy generation in non-equilibrium
thermodynamics are connected by the fluctuation theorem \cite{Ftheor2002},
which indicates that entropy increases are much more likely than entropy
reductions. The variational principles of non-equilibrium thermodynamics
have been previously reviewed in the literature \cite{MEP2006}.

The possibility of applying the entropy concept to macroscopic motions in
turbulence and other similar processes (rather than to thermodynamic
microstates) is the main interest of the present work. Pope \cite{Pope-79c}
suggested that in absence of any further information, the best way of
approximating pdfs in turbulent flows is maximisation of entropy of the pdf
constrained by available information about the pdf. Falkovich and Fouxon 
\cite{Falk2004} analysed turbulence spectra with the use of entropy defined
similar to the configurational entropy in the present work. Apparent
thermodynamics is quite successful in specifying properties of inverse
cascade in two-dimensional turbulence \cite{Som2001,2005}, since energy is
preserved in this cascade and there is no vortex stretching in two
dimensions.  Dupree \cite{Dupree1992} analysed two-dimensional turbulence and
introduced a definition of entropy which has two terms similar to
configurational and state terms in equation (\ref{Scs}). 
Three-dimensional
turbulence, however, has proven to be more difficult and less susceptible to
analysis based on the thermodynamic principles. Celani and Seminara \cite
{Celani2005} used DNS (direct numerical simulations) results to demonstrate
that the statistics of turbulent scalar transport differs from the
statistics expected in Gibbs equilibrium. Duplat and Villermaux \cite
{Duplat2008} considered random stirring of a scalar field and found that it
does not produce a field with maximal entropy. In all these works,
fluctuations are treated as bringing additional chaos and entropy of these
fluctuations is a positive quantity. Sancho and Llebot \cite{SANCHO1994},
however, suggest that the entropy associated with turbulent motion, which is
more ordered as compared to highly chaotic molecular motion, is negative.
This does not contradict the other publications since turbulent entropy is
defined in Ref. \cite{SANCHO1994} as the difference between the molecular
entropy in a turbulent flow and that in a notional state of the flow after
all turbulent fluctuations have been dissipated by viscosity. These examples
illustrate that different quantities can be introduced as entropy and be
very useful for analysis of non-equilibrium processes. However, when these
quantities are different from the conventional molecular entropy, it is
important to accurately define the quantity under consideration. In the
present work, we use the term apparent entropy to distinguish entropy-like
quantities from the molecular entropy. A similar distinction was drawn by
Gray-Weale and Attard \cite{Attard2007} who use the terms ``first entropy''
and ``second entropy'' to distinguish the quantities analogous to the
molecular and apparent entropies. Unlike molecular entropy, the apparent
entropy is not necessarily controlled by the laws of thermodynamics
(separately from molecular entropy), and its properties require a special
investigation.

\section{Entropy and mixing \label{sec_M0}}

\subsection{Entropy change by conservative mixing \label{sec_M2}}

In this subsection we consider how entropy is changed by mixing as simulated
by the major conventional mixing models. Mixing between particles can be
preferential, when particle properties affect the selection of particles for
mixing, or non-preferential. Although all mixing models we consider perform
mixing between particles locally in physical space $\mathbf{x}$,
non-preferential mixing models do not discriminate particles on the basis of
their properties $\mathbf{y}.$ In principle, preferential mixing allows for
additional adjustment of mixing models to match better the physical mixing
processes they simulate. Mixing preference can generally be expressed with
the use of the mixing connectivity function $0\leq \Psi _{pq}\leq 1$ defined
so that particles $p$ and $q$ can not mix if $\Psi _{pq}=0,$ particles $p$
and $q$ are most likely to be selected for mixing if $\Psi _{pq}=1.$ The
example problem is conventional homogeneous mixing of two initial states
with $y=0$\ and $y=1$ occurring with equal probability into the final state
of $y=1/2$.

The models commonly used in combustion applications include IEM
(Interactions by Exchange with the Mean, \cite{Dopazo75}), the Curl's \cite
{Curl-63a}\ and modified Curl's mixing models \cite{Janika79,LindFete03},
the EMST (Euclidean Minimal Spanning Tree) \cite{Subramaniam-P-97a} and the\
MMC (Multiple Mapping Conditioning) model \cite{KP03,book2011CK}. The two
last models represent stochastic versions of Mapping Closure (MC) \cite
{Chen-C-K-89} and, for the problem considered here, would perform similar to
MC. EMST introduces MC-type mixing through preferential mixing between
particles. MMC exercises preferential mixing between particles but only in
terms of the special reference variables that are added to the set of
physical coordinates $\mathbf{x}$; the selection of particles does not
depend on $\mathbf{y}$ during MMC mixing. Mixing in the conventional Curl's
model is non-preferential.

\begin{figure}[ht] 
\caption{Evolution of configurational entropy $S_c$ 
 in homogeneous turbulence simulated by 
 mapping closure (solid line) and 
 by modified Curl's model (dotted line)} 
\begin{center} 
\includegraphics[width=7cm]{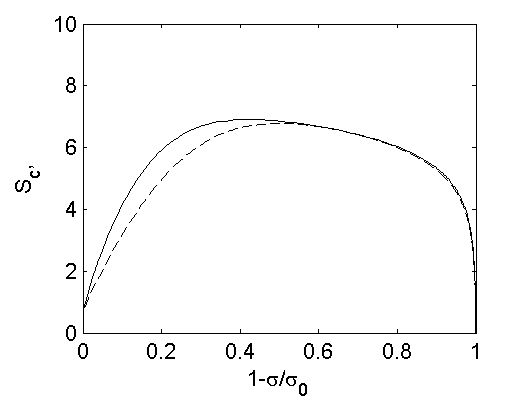}  
\label{fig_1}  
\end{center}
\end{figure}

Mixing affects both the configurational and state entropies. The change in
configurational entropy is considered first. If the initial pdf $f(y)$ has
Delta-functions, the IEM and the old Curl's model do no produce smooth pdf
distributions and are not suitable for this analysis. The mixing simulated
by modified Curl's model and MP results in smooth distributions for which
the configurational entropy is well defined. The Curl's model specifies
mixing of particle $p$ with another particle $q$ by formula 
\begin{equation}
\acute{y}_{p}=\frac{1+\eta }{2}y_{p}+\frac{1-\eta }{2}y_{q},
\label{mix-Curl}
\end{equation}
where the extent of mixing $\eta $ is constant for the old Curl's model and
random for the modified Curl's model. We use a uniform distribution of $\eta 
$ on the interval $[0,1]$. The calculated values of the configurational
entropy versus a time-like variable $1-\sigma /\sigma _{0}$\ where $\sigma
^{2}=\left\langle (y-\left\langle y\right\rangle )^{2}\right\rangle $ and $%
\sigma _{0}=\sigma (t_{0})$ are presented in Figure \ref{fig_1}. There is a
noticeable difference in entropies at the first stages of the mixing process
that becomes small in the final stages of mixing despite the fact that
MC-simulated pdf correctly approaches the Gaussian distribution while the
pdf simulated by the modified curls model does not. The pdf simulated by MC
is very close to the scalar pdf in real homogeneous turbulence \cite
{Pope-91a} and so should be the configurational entropy term shown in Figure 
\ref{fig_1}.

\begin{figure}[ht] 
\caption{Change in entropy of mixing $s(y)$ induced by Curl's mixing} 
\begin{center} 
\includegraphics[width=8cm]{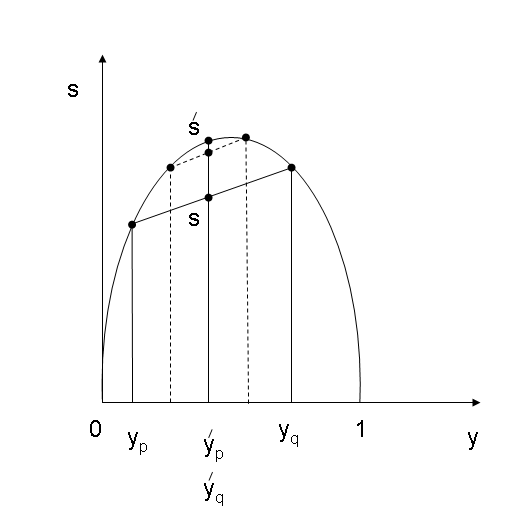}  
\label{fig_2}  
\end{center}
\end{figure}

We note that the configurational entropy can both increase and decrease in
simulations. This, of course, does not contradict to the second law of
thermodynamics as the second component --- the potential entropy --- must be
taken into account. For the case of ideal mixing the molecular entropy of
mixing is defined by 
\begin{equation}
s(y)=-\beta _{m}\left( y\ln (y)+(1-y)\ln (1-y)\right)   \label{s-mix}
\end{equation}
where the constant $\beta _{m}$ is introduced to account for Boltzmann
constant and scale different entropies consistently. The term $s(y)$ enters
equation (\ref{Scs}) as the potential entropy $S_{f}$ . We can show that
this quantity always increases when simulated by any mixing model producing
a non-negative approximation for the conditional scalar dissipation $%
N_{y}=\left\langle D(\nabla y)^{2}|y\right\rangle $. Indeed the pdf scalar
transport equation 
\[
\frac{\partial f(y)}{\partial t}=-\frac{\partial ^{2}N_{y}f(y)}{\partial
y^{2}}
\]
results in the following expression for the entropy change 
\[
\frac{dS_{f}}{dt}=\int_{0}^{1}\frac{\partial f(y)}{\partial t}s(y)dy=
\]
\begin{equation}
=-\int_{0}^{1}\frac{\partial ^{2}N_{y}f(y)}{\partial y^{2}}%
s(y)dy=-\int_{0}^{1}N_{y}f(y)\frac{d^{2}s(y)}{dy^{2}}dy\geq 0
\end{equation}
considering the fact that $d^{2}s/dy^{2}$ is negative for $s(y)$ defined by (%
\ref{s-mix}). The integral is evaluated here by parts while taking into
account that both $N_{y}f(y)$ and its derivative tend to zero at the
boundaries $y=0$ and $y=1$ \cite{KB99}. Figure \ref{fig_2} shows the curve $%
s(y)$ and demonstrates that after both complete and incomplete mixing of two
particles $p$ and $q$ the mean entropy of these particles $%
(s(y_{p})+(y_{q}))/2$ increases.

This example illustrates the following important point. The total entropy
can be treated as the conventional thermodynamic entropy with the potential
entropy $S_{f}$ representing chaos of molecules completely mixed to
molecular level and the configurational entropy $S_{c}$ representing the
chaos of turbulent fluctuations. In this case, however, the term $S_{f}$ is
proportional to the number of molecules and $\beta _{m}$ is so large that
the term $S_{c}$ is indistinguishable in the sum. The entropy $S=S_{c}+S_{f}$
does, of course, satisfy the second law of thermodynamics. It can be very
useful to consider the entropy of turbulent fluctuations but this quantity
needs to be examined separately from the molecular entropy.

\subsection{Entropy change by competitive mixing}

\textit{Abstract competition} studies the principles of competition in their
most generic form \cite{KlimIMCIC2010}. The purpose of this representation
may be seen to be similar to that of the Turing machine: making complex
behaviours susceptible to general analysis but not specifically simulating
any real devices or processes. Consider a complex system that has a large
number of autonomous elements engaged in competition with each other. The
evolution of a competitive system involves, in its general form,\ a process
of determining a winner and a loser for competition between any two elements
of the system. The properties of the loser are lost while the winner
duplicates its properties into the resource previously occupied by the
loser. The duplication process may involve random changes, which are
customarily called mutations irrespective of the physical nature of the
process. These mutations are predominantly negative or detrimental but can
occasionally deliver a positive outcome. It is easy to see that abstract
competition can be represented by a system of Pope particles, provided
conventional conservative mixing is replaced by competitive mixing as
discussed below. In context of computations, the competing elements or any
other notional autonomous objects are conventionally called particles
without implying a reference to physical particles of any kind. In the
present review ``elements'' and ``particles'' are used synonymously with
``elements'' primarily referring to competing components of general nature
and ``particles'' to their computational implementations.

Unlike conservative mixing, competitive mixing does not conserve the scalar
values and gives a priority to the particle in a mixing group that is
determined as the ``winner'' by duplicating its properties. The properties
that belong to the loser are lost. The binary relationship ``$\mathbf{y}_{p}$
stronger than $\mathbf{y}_{q}",$ or $\mathbf{y}_{p}\succ \mathbf{y}_{q},$
means that particle $p$ is the winner and particle $q$ is the loser in
competition of these particles. If $\mathbf{y}_{p}$ is stronger than $%
\mathbf{y}_{q}$ (i.e. $\mathbf{y}_{p}\succ \mathbf{y}_{q}$) then, by
definition, $\mathbf{y}_{q}$ is weaker than $\mathbf{y}_{p}$ (i.e. $\mathbf{y%
}_{q}\prec \mathbf{y}_{p}$)$.$ If $\mathbf{y}_{p}$\ and $\mathbf{y}_{q}$
have the same strength then write $\mathbf{y}_{p}\simeq \mathbf{y}_{q},$
while ``$\preccurlyeq $'' implies ``$\prec $'' or ``$\simeq $'' and ``$%
\succcurlyeq $'' implies ``$\succ $'' or ``$\simeq $''. The outcome of
competition is determined by the properties of the particles. Although more
complicated schemes can be considered, competitive mixing is introduced here
according to the following mixing rules \cite{K-PS2010} 
\begin{equation}
\mathbf{\acute{y}}_{p}=\left\{ 
\begin{array}{ccc}
\mathbf{y}_{q}+\mathbf{\zeta }, & \mathbf{y}_{p}\prec \mathbf{y}%
_{q},R_{pq}=-1 & (\func{loser}) \\ 
\left[ 
\begin{array}{c}
\mathbf{y}_{q}+\mathbf{\zeta } \\ 
\mathbf{y}_{p}
\end{array}
\right] , & \mathbf{y}_{p}\simeq \mathbf{y}_{q},R_{pq}=0 & (\func{draw}) \\ 
\mathbf{y}_{p}, & \mathbf{y}_{p}\succ \mathbf{y}_{q},R_{pq}=+1 & (\func{%
winner})
\end{array}
\right. ,  \label{R1}
\end{equation}
where $R_{pq}=-R_{qp}$ is the antisymmetric competition index and $\mathbf{%
\zeta }$ represents random mutations. Although this term is obviously
borrowed from biology, mutations used here represent random redistribution
of resources (i.e. particles) between different states and should not be
confused with genetic mutations. In the case of a draw, the two possible
outcomes are selected with equal probability. In principle, particles $p$
and $q$ can also be isolated from each other $\mathbf{y}_{p}\parallel 
\mathbf{y}_{q},$ which corresponds to $\Psi _{pq}=0$; in this case $R_{pq}$
does not need to be defined but still can be defined for these particles if
convenient. Competition can be illustrated by the following reaction between
the particles 
\begin{equation}
\mathbf{y}_{p}+\mathbf{y}_{q}+\func{energy}\rightarrow \mathbf{y}_{p}+%
\mathbf{y}_{p}^{\prime },\;\;\;\;\mathbf{y}_{p}\succ \mathbf{y}_{q}
\end{equation}
where $\mathbf{y}_{p}^{\prime }$ represents a mutated version of $\mathbf{y}%
_{p}$, $\mathbf{y}_{p}$ is stronger than $\mathbf{y}_{q}$ and the word
``energy'' indicates existence of an external source of energy.

The most simple example of competitive mixing is given by mixing of two
states the losing state of $y=0$ and the winning state of $y=1.$ The
properties of this mixing were investigated by Klimenko and Pope \cite
{KP2012}. This mixing\ can be considered as a model for turbulent premixed
combustion as well as for evolutionary invasion processes such as invasion
of more successful species into the area occupied by less successful
species. The model follows findings of Pope and Anand \cite{Pope-A-85} that
when reactions are fast the pdf of the reaction progress variable is
dominated by two states: unburned $y=0$ and burned $y=1.$ The governing
equation for this model is related to the KPP-Fisher equation \cite
{KPP1937,Fisher1937} and to the BML (Bray-Moss-Libby) model for premixed
turbulent combustion \cite{Bray-L-M-84a}. The equation is named after Fisher
who introduced it first and after Kolmogorov, Petrovsky and Piskunov who
investigated the mathematical properties of this class of equations. For
this case, the entropy equation (\ref{Scs}) takes the form 
\begin{equation}
S=\stackunder{S_{f}}{\underbrace{s_{1}\left\langle y\right\rangle }}\mathbf{-%
}\stackunder{S_{c}}{\underbrace{\mathbf{(}\left\langle y\right\rangle \ln
\left( \left\langle y\right\rangle \right) +(1-\left\langle y\right\rangle
)\ln ((1-\left\langle y\right\rangle ))}},
\end{equation}
where we put $s(0)=0$ and $s_{1}=s(1)$ without loss of generality and $%
\left\langle y\right\rangle $ is the average value of $y$ over the two
states $y=0$ and $y=1$. Although one may notice some similarity with
equation (\ref{s-mix}), the physical meaning of that equation is different.
As $\left\langle y\right\rangle $ increases due to mixing, configurational
entropy $S_{c}$ increases but then, as the winning state becomes more and
more dominant in the distribution, decreases. For combustion waves, this
correspond to converting reactants into the products and $S_{c}=0$ when the
reactions are complete and only products are present. The products have a
much higher value of entropy than the reactants (i.e. $s_{1}\gg 1)$ and this
ensures that the reactions are directed from reactants to the products.

The existence of thermodynamics driving chemical reactions towards their
equilibrium states is obvious. Since the same model based on Pope particles
can be used to simulate invasions \cite{KP2012}, there should be an apparent
thermodynamics, which can characterise the invasion and be similar to the
conventional thermodynamics of the chemical reactions mentioned above. While
noting the similarities between reactions and invasions, we should not
forget about the differences. Reactions in premixed combustion are directly
driven by conventional thermodynamics towards maximal molecular entropy (or
possibly minimal molecular Gibbs/Helmholtz free energy). In this case
apparent thermodynamics is directly linked to conventional thermodynamics.
Molecular entropy of successful species is not necessarily higher than (and
the Gibbs/Helmholtz free energy is not necessarily lower than) that of
unsuccessful species and the apparent and molecular quantities are not
directly linked. Competitive systems in this case must receive exergy from
outside to avoid the constraints imposed by conventional thermodynamics on
isolated systems. Positive values of the apparent entropy potential $s_{1}$
indicate the higher probability of presence of successful species in the
equilibrium mixture irrespective of the physical reasons that ensure this
success. While molecular thermodynamics explains the ``success'' of products
over the reactants, it is not likely to offer a universal justification for
the success of some competing elements over the others. Apparent
thermodynamics recognises the obvious: nature has a greater affinity towards
some states or competitive elements as compared to the other states or
elements, irrespective whether we have an explanation for this affinity or
not. Apparent thermodynamics is not fully reducible to molecular
thermodynamics in the same way as molecular thermodynamics is not fully
reducible to the laws of conventional and quantum mechanics.

The invasion process is redistribution of the available resources in favor
of the successful species. Mutations represent randomness in this process
and may or may not be related to genetic mutations. The relativistic nature
of the competitiveness should be stressed. Weaker species are perfectly
stable and become weak only in presence of stronger species in the same way
as reactants disappear only when their transformation into products is
allowed. We now proceed further to introduce a special thermodynamics that
can characterise competitive systems.

\section{Ordering, ranking and entropy}

\subsection{Ranking in competitive systems}

Ranking of particles or elements in competition reflects how well a particle
performs relative to the other particles. We distinguish the following
rankings:

\begin{enumerate}
\item  \textbf{Two-particle ranking} is the index function $R_{pq}=R(\mathbf{%
y}_{p},\mathbf{y}_{q})$ that determines the winner and the loser in
competition of $p$ and $q$ as shown in equation (\ref{R1})

\item  \textbf{Absolute ranking} is a function $r_{\#}(\mathbf{y}_{p})$ that
determines the outcomes of the competition by 
\begin{equation}
r_{\#}(\mathbf{y}_{p})\leq r_{\#}(\mathbf{y}_{q})\;\;\Leftrightarrow \;\;%
\mathbf{y}_{p}\preccurlyeq \mathbf{y}_{q}  \label{abs_r}
\end{equation}
that is $r_{\#}(\mathbf{y}_{p})\leq r_{\#}(\mathbf{y}_{q})$ when and only
when $q$ is not a loser in competition with $p.$ Introduction of absolute
ranking requires transitivity and is subject to additional conditions as
discussed in the following subsections.

\item  \textbf{Relative ranking} $r(\mathbf{y}_{p}\mathbf{,[}f])$ is ranking
of a particle $\mathbf{y}_{p}$ relative to a given distribution $f(\mathbf{y}%
)$%
\begin{equation}
r(\mathbf{y}_{p}\mathbf{,[}f])=\stackunder{\infty }{\int }R(\mathbf{y}_{p},%
\mathbf{y}^{\prime })f(\mathbf{y}^{\prime })d\mathbf{y}^{\prime },
\label{rank-1}
\end{equation}
which indicates how competitive particle $p$ is with respect to distribution 
$f(\mathbf{y}).$ The function $-r(\mathbf{y}^{\prime }\mathbf{,[}f])$ can
also be interpreted as ranking of distribution $f(\mathbf{y})$ relative to
the location $\mathbf{y}^{\prime }$.

\item  \textbf{Co-ranking} $\bar{R}([f_{1}],[f_{2}])$ is relative ranking of
two distributions $f_{1}(\mathbf{y})$ and $f_{2}(\mathbf{y})$ defined by 
\begin{equation}
\bar{R}([f_{1}],[f_{2}])=\stackunder{\infty }{\int \int }R(\mathbf{y},%
\mathbf{y}^{\prime })f_{1}(\mathbf{y})f_{2}(\mathbf{y}^{\prime })d\mathbf{y}d%
\mathbf{y}^{\prime }  \label{rank-ee}
\end{equation}
and indicating competitive strength of these distributions with respect to
each other. Note that co-ranking is anti-symmetric: $\bar{R}%
([f_{1}],[f_{2}])=-\bar{R}([f_{2}],[f_{1}])$ and $\bar{R}([f_{1}],[f_{1}])=0$%
. \ If $\bar{R}([f_{1}],[f_{2}])>0,$ we may write $[f_{1}]\succ \lbrack
f_{2}]$ and say that the distribution $f_{1}$ is stronger than $f_{2}$ or,
if $\bar{R}([f_{1}],[f_{2}])=0,$ we may write $[f_{1}]\simeq \lbrack f_{2}]$
and say that the both distributions have the same strength.
\end{enumerate}

In Appendix \ref{sA2}, these definitions are generalised for preferential
mixing. The competitive binary relation, which is considered here, orders
competing elements and is connected to their ranking. This and the example
given in the previous subsection indicate the existence of a link between
ranking and the entropy potential. For example, if the absolute ranking is
introduced, then entropy potential $s$ can be deemed to be a function of $%
r_{\#}$ and higher ranking is expected to correspond to higher $s$. Higher
ranking and higher entropy potential recognise a greater affinity of nature
towards these states, while the physical reasons responsible for this
affinity may differ. For example, more competitive states may correspond to
higher molecular entropy or lower molecular Gibbs/Helmholtz free energy ---
in these cases the apparent and conventional thermodynamics are directly
linked. More competitive states may also correspond to higher production
rates of molecular entropy --- apparent thermodynamics can reflect the MEP
principle or, in fact, any other related variational principle. Following
the traditions of classical thermodynamics, we generally leave the exact
physical mechanism of competitiveness of the elements outside our
consideration but accept that some states are more competitive than others
and proceed to investigate the consequences. Competitive systems are, of
course, compliant with molecular thermodynamics but, at the same time, they
represent open systems and the apparent quantity $s$ is not necessarily
linked to the molecular entropy or Gibbs/Helmholtz free energy. The
connection between ordering, ranking and entropy has, as reviewed below,
cross-disciplinary significance. This connection is further explored in the
following sections, where we draw an important distinction between
transitive and intransitive competitions.

\subsection{Ranking and fitness}

The concept of fitness has similarities with competitive ranking, although
these concepts have differences. Ranking differs from fitness in the same
way that competition differs from criteria-based selection. A high-ranking
particle can perform poorly when even higher ranking competitors are present
while a low-ranking particle may survive if it does not have to compete
against particles with higher ranks. Traditional fitness reflects adaptation
to the environment and has an absolute value, which typically indicates the
percentage of surviving offspring, while ranking reflects a direct
competition between elements and is inherently relativistic. If the
differences between adaptation and competition are overlooked and fitness is
defined as a general indicator of the overall ability to survive, the
absolute ranking (provided, of course, the absolute ranking exists) can be
identified with fitness.

We should mention Eigen's quasispecies models \cite{Eigen1971}, which can
also duplicate and mutate elements. The essence of the current approach is
the direct competition between the elements comprising the system while the
elements of the Eigen model do not compete directly against each other but
utilise a common restricted resource with efficiency determined by the
fitness of the elements. The behaviour of competing elements changes
dramatically depending on which competitors are currently present, while the
relationships between different elements expressed by (\ref{R1}) can be very
complex. Competition makes a very sharp judgment: a loss by a small margin
is still a loss. As in the Eigen model, the competition may be powered by an
external source of exergy but, otherwise, the abstract competition, which we
consider here, is much more similar to conventional mixing than to
self-replication taking place in the Eigen model.

\subsection{Ranking and adiabatic accessibility}

A number of publications \cite{Giles1964,Cooper1967,Buch1975,Landsberg1978}
has been dedicated to the goal of constructing thermodynamics based on the
principle of \textit{adiabatic accessibility} \cite{Cara1909}. A notable
success has been achieved by Lieb and Yngvason \cite{EntOrd2003}, who
reviewed the previous attempts and demonstrated that this goal can be
achieved in a rigorous and unambiguous manner\footnote{ A similar approach, 
albeit based on a different quantity  -- 
{\it adiabatic availability}, has been previously developed by Gyftopoulos and Beretta
\cite{Beretta1991}}. A popular presentation of
these results is given by Thess \cite{Thess2011}. Adiabatic accessibility is
a binary relationship that indicates the possibility or impossibility of
reaching one state from the other by a reversible or irreversible adiabatic
process. This binary relation and the rest of conventional thermodynamics
are fundamentally transitive. Adiabatic accessibility is required to comply
with a number of axioms including transitivity and allows for the
introduction of empirical entropy that remains the same in reversible
processes and increases in irreversible processes \cite{Buch1975}. Empirical
entropy is not unique: any strictly monotonic continuous function of the
empirical entropy is its equivalent. One of these functions, however, is
thermodynamically extensive and represents the thermodynamic entropy. If we
use the current notations, then $\mathbf{y}_{p}\preccurlyeq \mathbf{y}_{q}$
indicates that the state $\mathbf{y}_{q}$ is adiabatically accessible from
the state $\mathbf{y}_{p}$ by a reversible adiabatic process when $\mathbf{y}%
_{p}\simeq \mathbf{y}_{q}$ or by an irreversible adiabatic process when $%
\mathbf{y}_{p}\prec \mathbf{y}_{q}$. The empirical entropy is analogous to
absolute ranking $r_{\#}(\mathbf{y)}$ while $s$ is related to thermodynamic
entropy. The function $s=s(r_{\#})$ is monotonic and represents equivalent
ranking but $s$ is also constrained by the properties of mutations. The
analogy with adiabatic accessibility is transparent.

\subsection{Ranking and economic utility}

The introduction of an absolute ranking for transitive ordering is subject
to conditions of the Debreu theorem \cite{Debreu1954} (see Appendix \ref{sA1}%
), which was originally formulated in context of economic science, where
absolute ranking of consumer preferences has been repeatedly studied under
the name of ``utility'' (see review by Mehta \cite{Mehta1999}). Utility
specifies the competitive property of some goods and services to satisfy the
needs of consumers as compared to that of other goods and services.

It is most useful to learn that similar methods have been under development
in theoretical physics and mathematical economics for more than half a
century without any knowledge or interaction between these fields. The
similarity between introducing economic utility and physical entropy was
noticed first by Candeal et. al. \cite{Mehta2001}, who called the similarity
``astonishing''. While Candeal et. al. \cite{Mehta2001} proceeded further to
compare the formal conditions of the main theorems, the principal question
of the physical reasons behind this similarity remained unanswered. If
abstract competition is relevant to both thermodynamical entropy and
consumer utility, this may serve as the missing physical link between the
fields. Although abstract competition is a generic framework, which is not
intended to simulate any specific economic conditions, the following
consideration indicates that, indeed, consumer behaviour might be related to
abstract competition and there probably should be a kind of economic entropy
associated with utility.

The traditional economic consumer has to solve a conditional extremum
problem while going shopping --- the problem of maximising the utility of
his consumption bundle under given budgetary constraints. A less
mathematically savvy consumer, who behaves according to the competition
principles considered here, simply compares his existing bundle $\mathbf{y}%
_{p}$ with another offered bundle $\mathbf{y}_{q}$ and, if he likes $\mathbf{%
y}_{p}$ more than $\mathbf{y}_{q}$ (i.e. $\mathbf{y}_{p}\succ \mathbf{y}_{q}$%
), keeps the existing bundle $\mathbf{y}_{p}.$ The consumer, however, can
like the new offering more than the old one (i.e. $\mathbf{y}_{q}\succ 
\mathbf{y}_{p}$), then in this case $\mathbf{y}_{p}$ is replaced by $\mathbf{%
y}_{q}$. Economists may say that the consumer reveals his preference of $%
\mathbf{y}_{q}$ over $\mathbf{y}_{p}$. The analogy can be extended to
involve mutations: if $\mathbf{y}_{q}\succ \mathbf{y}_{p},$ the consumer may
not get exactly what he/she wants or expects (i.e. $\mathbf{y}_{q}$ --- one
may recall inaccurate advertising or incomplete information about the
products) but a modified version of the bundle $\mathbf{\acute{y}}_{p}=%
\mathbf{y}_{q}+\mathbf{\zeta .}$ It is most likely that the consumer would
not like these modifications $\mathbf{y}_{q}+\mathbf{\zeta }\prec \mathbf{y}%
_{q}$ as, indeed, the mutations tend to be predominantly negative.

\section{Thermodynamics of transitive competition \label{sec_T0}}

Competition is deemed transitive when for any selected particles $p,$ $q$
and $r,$ 
\begin{equation}
\mathbf{y}_{p}\preccurlyeq \mathbf{y}_{q}\func{ and }\mathbf{y}%
_{q}\preccurlyeq \mathbf{y}_{r} \Rightarrow \mathbf{y}_{p}\preccurlyeq 
\mathbf{y}_{r}  \label{trans1}
\end{equation}
that is the relations $\mathbf{y}_{p}\preccurlyeq \mathbf{y}_{q}$ and $%
\mathbf{y}_{q}\preccurlyeq \mathbf{y}_{r}$ demand that $\mathbf{y}%
_{p}\preccurlyeq \mathbf{y}_{r}$. Transitive binary relationships of this
kind can be referred to as order or a preorder. Subject to the conditions of
the Debreu theorem \cite{Debreu1954}, transitive competition allows for
introduction of the absolute ranking $r_{\#}=r_{\#}(\mathbf{y})$ defined by (%
\ref{abs_r}).

The absolute ranking is related to the entropy potential: higher ranking of $%
\mathbf{y}_{q}$ corresponds to higher probability of this state and
consequently to higher entropy potential $s(r_{\#}(\mathbf{y}_{p}))$. In the
example of premixed combustion model of the previous section, the absolute
ranking can be selected so that the states of $y=0$ and $y=1$ correspond to $%
r_{\#}=0$ and $r_{\#}=1$ respectively. so that higher rank corresponds to a
stronger particle. If mixing is non-preferential, it is sufficient to
consider a single property $r_{\#}(\mathbf{y})$ that can be simply denoted
by $y$. It should be noted that absolute ranking is not unique and any
monotonically increasing function of $r_{\#}$ represents an equivalent
ranking. Similarities with existing approaches are explored in the following
subsections.

\subsection{Gibbs mutations \label{sec_T1}}

If mixing is non-preferential and the competition is transitive, the
outcomes of the competition are determined only by the absolute ranking $%
r_{\#}(\mathbf{y})$. For the sake of simplicity, we can assume that $\mathbf{%
y}$ is a scalar denoted by $y$ since, otherwise, we can simply select
ranking $r_{\#}$ as $y$. Knowledge of $y_{p}$ and $y_{q}$ is sufficient to
determine the winner in competition between particles $p$ and $q$. We imply
that higher values of $y$ correspond to higher ranking, hence $%
y_{1}\preccurlyeq y_{2}$ is the same as $y_{1}\leq y_{2}$.

Thermodynamic relations become most transparent for a certain class of
mutations that satisfy some Markovian restrictions and are named \textit{%
Gibbs} \textit{mutations. }As discussed in Appendix \ref{sAm}, we broadly
follow the ideas of introducing thermodynamically consistent Gibbs measures
for Markov fields and graphs \cite{GibbsMeasure80}. Gibbs mutations are
non-positive and for the case considered here take the form 
\begin{equation}
f_{\zeta }(y,y^{\circ })=\left\{ 
\begin{array}{cc}
\frac{f_{0}(y)}{F_{0}(y^{\circ })}, & y\leq y^{\circ } \\ 
0, & y>y^{\circ }
\end{array}
\right. ,  \label{mut-f0}
\end{equation}
where $F_{0}(y^{\circ })$ is the normalisation constant depending on $%
y^{\circ }$ and $f_{0}(y)$ is the equilibrium distribution (that is
according to the H-theorem \ref{thH1}, the distributions $f(y\mathbf{)}$
converges to the same function $f_{0}(y\mathbf{)}$ that is used in the
definition of $f_{\zeta }.$ Equation (\ref{mut-f0}) is consistent with (\ref
{Gibbs3a})\ and also with a more general definition of Gibbs mutations by (%
\ref{mut-tc}).

Absolute ranking is generally not unique since any monotonically increasing
function of $r_{\#}$ represents an equivalent ranking.\ We relate absolute
ranking to entropy $s=s(r_{\#})$\ potential\ and can use this entropy for
ranking purposes. The equation $f_{0}\sim \exp (s)$ links $s$ to the
mutation intensity and makes this entropy-related definition of ranking
unique. The a priori statistical weight $A(y\mathbf{)}$ can account for
different phase volumes of different states and in many cases can formally
be set to unity without affecting the evolution of the system. However, the
physical interpretation of $A(y\mathbf{)}$ can be linked to the probability
of particle distribution under conditions when the competition is switched
off. We should note that the exponential form for distribution of mutations
was previously suggested and used in genetic theory \cite{Ohta1977},
although we do not have any specific intention here to match the properties
of genetic mutations and are interested in a general consideration of
competing systems. Theorem \ref{thH1}, which is proved in Appendix \ref{sA5}
and represents the principal step for introducing competitive
thermodynamics, is a competitive analog of the Boltzmann H-theorem.
According to this theorem, the entropy $S$ monotonically increases until it
reaches its maximal value and at this point the distribution $f(y)$ reaches
its equilibrium $f_{0}(y)$. The particle with the highest rank --- the
leading particle, which denoted here by $y_{\ast }$ --- remains at the same
location since it can not lose competition to a lower-ranker and at the same
time can not be overtaken by another particle due to absence of positive
mutations.

The H-theorem also indicates that a detailed equilibrium is reached in the
equilibrium state. In this state the overall entropy is maximal and, if the
system is divided into subsystems, say $I$ and $J$ (see Appendix \ref{sA2}),
their competitive potentials must be the same $\chi _{I}=\chi _{J},$
otherwise the entropy can be increased by transferring particles from the
subsystem with lower $\chi $ to the subsystem with higher $\chi $ (note that
according to equation (\ref{mu}) $\chi _{I}=\ln (Z_{I}/a_{I})-1$ for any $I$
). Due to the detailed balance in equilibrium, the competitive connection
between any two locations, say $y_{1}$ and $y_{2},$ can be severed without
affecting the equilibrium state (terminating both the competition and the
exchange by mutations) as long as these locations remain connected through
other locations. Competitive systems with Gibbs mutations are thus most
stable and stability is an important factor constraining the existence of
any realistic system.

\subsection{Infrequently positive near-Gibbs mutations}

The existence of positive mutations is an important factor affecting the
evolution of competitive systems which can not be overlooked or neglected
even if these mutations are small and infrequent.\ Positive mutations are
deemed to be relatively rare and negative mutations remain dominant. The
absolute ranking $r_{\ast }=r_{\#}(y_{\ast })$ of the leading particle still
can not decrease but can increase as there is a small but still positive
probability of mutations that result in a particle overtaking the leader and
becoming the leading particle. The state with distribution $f_{0}(y,y_{\ast
})$ and a fixed $y_{\ast }$ should be referred to as a \textit{%
quasi-equilibrium} state since the distribution may shift towards higher
ranks whenever the leading particle is overtaken. The leading particle $%
y_{\ast }$ is occasionally overtaken by another particle $y_{\ast }^{\prime
} $ due to a positive mutation in the leading group so that $r_{\ast
}^{\prime }=r_{\#}(y_{\ast }^{\prime })>r_{\ast }=r_{\#}(y_{\ast })$. The
distribution function remains almost without change $f_{0}(y,y_{\ast })$ but
it is now different from the new equilibrium distribution $f_{0}(y,y_{\ast
}^{\prime }).$\ According to the H-theorem, $f_{0}(y,y_{\ast })$ should
evolve towards $f_{0}(y,y_{\ast }^{\prime })$ as entropy increases. The
current system can be treated as a combination of two subsystems with the
domains $\frak{D}_{\ast }$ and $\frak{D}_{\Delta }$ corresponding to the
intervals $r_{\#}(y)\leq r_{\ast }$ and $r_{\ast }<r_{\#}(y)\leq r_{\ast
}^{\prime }$. The particles move between the subsystems towards higher
values of competitive potential $\chi ,$ that is from $\frak{D}_{\ast }$ to $%
\frak{D}_{\Delta }$ until the equilibrium between the subsystems is reached.
The overall distribution shifts from $f_{0}(y,y_{\ast })$ into the more
competitive state of $f_{0}(y,y_{\ast }^{\prime })$.

The behaviour of competitive systems with infrequently positive mutations is
still consistent with the introduced thermodynamics and results in
increasing total entropy $S(t)$ and competitive potential $\chi _{0}(t)$.
Equilibration from arbitrary initial conditions occurs in two steps: rapid
relaxation into quasi-equilibrium $f_{0}(y,y_{\ast })$ with fixed $y_{\ast }$
and 2) gradual increase of the system ranking $r_{\ast }(t)=r_{\#}(y_{\ast
}(t))$ in time. The second process is, rigorously, still not at equilibrium
but as long as the probability and the magnitude of positive mutations are
small, the current distribution $f$ remains close to the equilibrium
distribution $f_{0}(y,y_{\ast }(t))$ that depends on the currently attained
ranking $r_{\ast }(t)$ of the leading particle $y_{\ast }(t)$. Hence, $f$ is
given by equation (\ref{eq-f0}) with 
\begin{equation}
A(y,y_{\ast }(t))=A_{0}(y)H(y_{\ast }(t)-y),
\end{equation}
where $H$ is the Heaviside function and the partition function (\ref{eq-Z})
becomes time-dependent 
\begin{equation}
Z(t)=\int_{-\infty }^{y_{\ast }(t)}A_{0}(y)\exp \left( s(y)\right) dy,
\end{equation}
Competition resulting in gradual overall increase in absolute ranking and in
competitive potential $\chi _{0}(t)=\ln (Z(t))-1$ is called \textit{%
competitive escalation}. If and when the leader reaches its maximal possible
rank, the system enters the state of \textit{global equilibrium}, which can
be altered only by external forces.

\subsection{General infrequently positive mutations}

Although Gibbs mutations represent a reasonable and general approximation
for the randomness present in competitions, mutations may deviate from this
approximation. For example, this may happen if the accessible space becomes
dependent on the location of the leading particle. In terms of the a priori
statistical weight this can be expressed as $A=A(\mathbf{y,y}_{\ast })$. \
The competition considered in this subsection is transitive with absolute
ranking $r_{\#}(\mathbf{y})$. If the competition is transitive and the
position of the leading particle is fixed, the process according to the
second convergence theorem presented in Appendix \ref{sA5} still converges
to its equilibrium state $f_{0}$ with maximal entropy $S$ although
convergence is not necessarily monotonic and the detailed balance is not
necessarily achieved in the steady state. The shape of the equilibrium
distribution is dependent on the position of the leading particle.

The position of the leading particle either remains fixed if the mutations
are non-positive (and no particle can overtake or challenge the leader), or
the leading particle $\mathbf{y}_{\ast }(t)$ escalates towards higher ranks: 
$\partial r_{\#}(\mathbf{y}_{\ast }(t))/\partial t>0$ if mutations are
infrequently positive. Small and infrequent positive mutations should not
affect a distribution that remains close to the equilibrium $f\approx f_{0}(%
\mathbf{y,y}_{\ast }(t)).$\ If the parameters of the competition do not
change with $\mathbf{y}_{\ast }$ the shape of the function remains the same $%
f\approx f_{0}(\mathbf{y-y}_{\ast }(t))$ while location of the function
shifts towards higher ranks. Hence, the behaviour considered here is very
similar to the case with near-Gibbs mutations: rapid relaxation of the
distribution into a quasi-equilibrium state $f_{0}(\mathbf{y,y}_{\ast })$
and then a gradual escalation of the distribution towards higher ranks.

Overall, a competitive system with general infrequently positive mutations
and transitive competition behaves qualitatively similarly to the case of
Gibbs mutations, but the analogy with conventional thermodynamics weakens.

\subsection{Competition and principles of non-equilibrium thermodynamics}

The trend of moving towards higher ranking is consistent with the introduced
competitive\ thermodynamics since the total entropy increases in this
process. Although Prigogine's theorem of minimal entropy production \cite
{PrigGlan} is generally not applicable to this process, we note that the
behavior of the system with infrequently positive mutations is qualitatively
consistent with the theorem. Indeed, if the initial distribution $%
f(y_{1},y_{\ast })$ is far from equilibrium, the system rapidly approaches
the equilibrium distribution $f_{0}(y_{1},y_{\ast })$ and the entropy
production significantly decreases as the distribution becomes close to $%
f_{0}(y_{1},y_{\ast })$. If positive mutations are present in the system,
then the entropy production continues at a small rate as the distribution
gradually moves towards larger values of $r_{\#}$ and $s$. Application of
the MEP principle to complex systems can be subject to different
interpretations \cite{MAP2005,MEP2006,SWENSON2009}. One of the possibilities
is applying MEP to apparent entropy. We may expect that nature favours
quasi-steady processes with the fastest possible rate of increase (or
minimal rate of decrease) of competitiveness (and, consequently the apparent
entropy) among all possibilities available to the system. This seems
plausible as the systems achieving the most competitive states are expected
to be the winners in the competition. The possibility of applying MEP to the
rate of production of the molecular entropy is not clear as
self-organisation, which is present in competitive systems and discussed in
the following sections, tends to reduce the intensity of competition and, as
a result, reduce the production rate of conventional entropy (while increase
in numbers of successful species should indeed increase consumption of the
resources and the molecular entropy production).

In our consideration, information has eternal properties: once some
combination of particle properties is achieved, it can exist forever unless
destroyed in competition. We may also consider the case when particles have
a finite life span, as quasispecies have in the Eigen model \cite{Eigen1971}%
. The particles are to be terminated (and regenerated with random
properties) with some small probability after a lengthy characteristic time $%
\tau _{e},$ which we may call the erosion time as this process is likely to
be associated with increase in molecular entropy. In this case, the leading
particle may eventually disappear resulting in a weak drift towards lower
ranks. The presence of some randomness in determining the winner and the
loser would have a similar effect.

The mutations considered here are predominantly negative. The physical
reason behind the rarity of positive mutations is not an inherent propensity
of the mutations for weakness: the mutations are random and do not have any
``purpose'', positive or negative. Mutations, however, tend to be negative
since there are many more effective micro-states $\Gamma (y)$ at the lower
ranks than at the higher ranks. Hence, a purely random mutation is much more
likely to step down than to step up in ranks. Here, we consider scalar $y$
for the sake of simplicity although the consideration is also suitable for
vector states $\mathbf{y.}$ It is possible to introduce the a priori entropy 
$\hat{s}(y),$ which is linked to the number of micro-states at given state $%
y $ and defined by the Boltzmann relation $\hat{s}(y)=\ln (\Gamma (y)).$
Here, $\Gamma (y)$ represents \textit{a priori} probability, which is the
nominal steady-state probability distribution in absence of competition. As
previously noted, $A(y)$ can be justifiably identified with $\Gamma (y)$ but 
$A(y)$ also may be selected in different ways without loss of generality.
For presentation of results in this section, it is convenient to simply put $%
A(y)\sim 1$.

\ Although the entropy $\hat{s}$ may often be implied while referring to the
entropy of evolutionary systems, the a priori entropy $\hat{s}$ is different
from and should not be confused with the entropy potential $s.$ The entropy
potential is related to competition and is larger at higher ranks while $%
\hat{s}$ is not related to competition and is larger at lower ranks.
Although $\hat{s}$ does not enter the definition (\ref{Scs}), the effect of
the a priori entropy is present in competitive thermodynamics. First we note
that there is an equilibrium distribution $\hat{f}_{0}(y)\sim \exp (\hat{s}%
(y)),$ which generally should be approached when competition is switched
off. In fact $\hat{f}_{0}$ can not be approached in most cases due to the
enormous capacity of $\Gamma ;$ hence $\hat{f}$ has to stay far from the
equilibrium $\hat{f}_{0}$. The a priori entropy, however, still influences
competition through non-equilibrium mechanisms as expressed by the
fluctuation theorem \cite{Ftheor2002}. Appendix \ref{sAm} demonstrates that
relative frequency of positive mutations is constrained by the fluctuation
theorem so that 
\begin{equation}
f_{\zeta }(\zeta )=f_{\zeta }(-\zeta )\exp \left( -\hat{\beta}\zeta \right)
,\;\;-\frac{d\hat{s}}{dy}=\hat{\beta}  \label{eqf-FT}
\end{equation}
Assuming that $\hat{\beta}>0\;$and $\zeta >0,$ one can note that the
distribution $f_{\zeta }$ has a steeper exponent on the positive side so
that \ large $\hat{\beta}$ enforces infrequency of the positive mutations.
The absence of positive mutations corresponds to infinitely large $\hat{\beta%
}$. Note that the condition $\hat{\beta}\gg 1$ limits the possible range of $%
y$ by physical restrictions imposed on $\Gamma $.

\subsection{Numerical simulations of competitive escalation}

In the special case of $A=1$ and linear dependence of $s$ on $y,$ that is $%
s=\beta y+\func{const},$ the distributions of mutations become exponential.
The value of $\beta $ may be seen as resembling conventional inverse
temperature, being inversely proportional to the intensity of fluctuations.
Equations (\ref{eq-f0}), (\ref{mut-f0}) and (\ref{eqf-FT}) yield a
double-exponential distribution 
\begin{equation}
f_{\zeta }(y,y^{\circ })=\left\{ 
\begin{array}{cc}
b_{0}\exp \left( b_{1}(y-y^{\circ })\right) , & y\leq y^{\circ } \\ 
b_{0}\exp \left( b_{2}(y^{\circ }-y)\right) , & y\geq y^{\circ }
\end{array}
\right.  \label{D-exp}
\end{equation}
where $b_{0}$ is determined by the normalisation condition 
\begin{equation}
\frac{1}{b_{0}}=\frac{1}{b_{1}}+\frac{1}{b_{2}}
\end{equation}
and 
\begin{equation}
b_{1}=\beta =\frac{ds}{dy},\;\;b_{2}=b_{1}+\hat{\beta},\;\;\hat{\beta}=-%
\frac{d\hat{s}}{dy}
\end{equation}
Note that both values $\beta $ and $\hat{\beta}$ are positive. Another
distribution, which is used in simulations presented here, is the shifted
exponential distribution given by 
\begin{equation}
\ f_{\zeta }(y,y^{\circ })=\left\{ 
\begin{array}{cc}
b_{1}\exp (b_{1}(y-y^{\circ }-\zeta _{0})), & y\leq y^{\circ }+\zeta _{0} \\ 
0, & y>y^{\circ }+\zeta _{0}
\end{array}
\right.  \label{eqf}
\end{equation}
Here, a small value $\zeta _{0}\geq 0$ accounts for rare positive mutations.
Without loss of generality, one may put $b_{1}=1$ and $y=s$.

\begin{figure}[ht] 
\caption{Quasi-equilibrium distributions for infrequently (4\%) positive mutations 
(dashed line -- double-exponential mutations;  
 dash-dotted line -- shifted exponential mutations) and 
 equilibrium distributions for non-positive mutations 
(solid line -- numerical simulations; circles -- exponential distribution ) } 
\begin{center} 
\includegraphics[width=12cm]{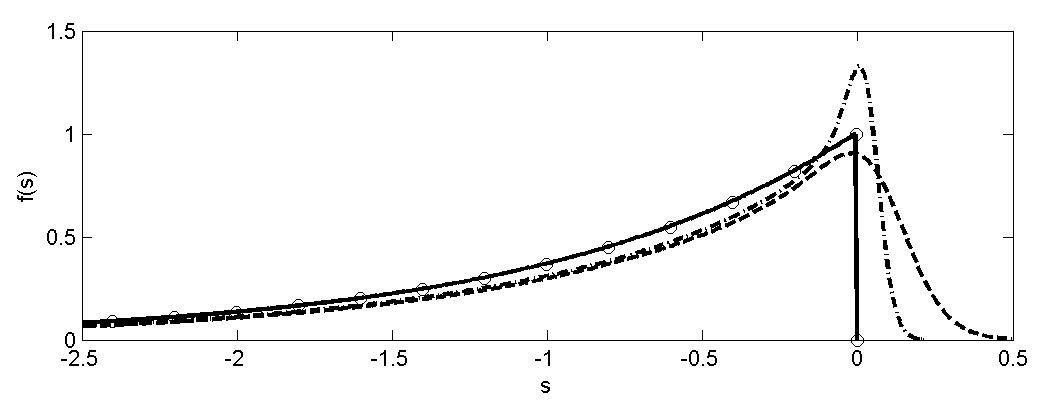}  
\label{fig_3}  
\end{center}
\end{figure}

The mutations are non-positive when $b_{2}=\infty $ in (\ref{D-exp}) or $%
\zeta _{0}=0$ in (\ref{eqf}). The mutations, however, become infrequently
positive when $b_{2}$ is large or $\zeta _{0}$ is small. The exact
analytical solution $f_{0}=\exp (y-y_{\ast })$ for non-positive mutations is
shown in Figure \ref{fig_3} by\ the solid line. The other lines show the
distribution $f(y,y_{\ast })$ for the process of competitive escalation with 
$4\%$ of positive mutations (i.e. $1/b_{2}=\zeta _{0}=0.04$). The
simulations are performed with 1000000 Pope particles. Mutations (\ref{D-exp}%
) and (\ref{eqf}) correspond to the dashed and dotted lines.\ The formula
determining the rate of competitive escalation, which was derived in Ref. 
\cite{K-PS2010}, can be written as 
\begin{equation}
\frac{dy_{\ast }}{dt}=\frac{c_{0}}{\Delta t}\left\langle \zeta H(\zeta
)\right\rangle \approx \frac{c_{0}}{\Delta t}\frac{\beta }{\hat{\beta}^{2}},
\label{dydt}
\end{equation}
where $H$ is the Heaviside function, $\Delta t$ is the time step and $c_{0}$
is a constant depending on the distribution of mutations. The approximate
evaluation of $dy_{\ast }/dt$ is performed in (\ref{dydt}) for $b_{2}\gg
b_{1}$ in the double-exponential distribution (\ref{D-exp}), indicating that
the effective selection rate of the competition process is given by $d\hat{s}%
_{\ast }/dt\sim -\beta /\hat{\beta}.$ The constant is $c_{0}\approx 5.4$ for
(\ref{D-exp}), $c_{0}\approx 3.2$ for (\ref{eqf}) and $c_{0}\approx 3.7$ for
the uniform distribution of mutations considered in Ref. \cite{K-PS2010}.

\section{Thermodynamics of intransitive competition.}

Competition is intransitive if at least one intransitive triplet 
\begin{equation}
\mathbf{y}_{p}\preccurlyeq \mathbf{y}_{q}\preccurlyeq \mathbf{y}_{r}\prec 
\mathbf{y}_{p}\   \label{trans2}
\end{equation}
exists in the system. Generally, a consistent absolute ranking can not be
introduced for intransitive competition, but ranking can often be assigned
to subdomains if the competition is transitive within these subdomains. It
should be noted, however, that the rankings assigned in different subdomains
would result in multi-valued functions and can not be made fully consistent
with each other when the competition is intransitive (see the example in
Figure \ref{fig_4}). Relative rankings are valid for all competitive systems
irrespective of their transitivity. The problem of introducing ranking in
intransitive tournaments has been treated in a number of relatively recent
publications \cite{Rank2004,Rank2005}. Our choice of ranking specified by (%
\ref{abs_r})-(\ref{rank-ee}) is based on consistency with the evolution
induced by competitive mixing. The term ``ordering'' conventionally refers
to transitive orders while intransitive binary relationships may be called
``preferences'' or\ ``tournaments''. The term ``tournament'' seems to have
become common in recent publications \cite{BinaryOp2010}, unfortunately,
this term is likely to be confusing in the context of the present work.

\subsection{Intransitivity and its physical reasons}

Although the fact that intransitivity may appear as the result of
superimposing several perfectly transitive rules has been known since the
days of French revolution as the Condorcet paradox (this paradox was noted
first by outstanding mathematician, philosopher and humanist marquis de
Condorcet \cite{Cond1785}), intransitivities were viewed for a long time as
something illogical or undesirable \cite{Intrans1964}. For example, if
someone prefers A to B, B to C and C to A, can we see this individual as
behaving reasonably? According to the famous Arrow theorem \cite{Arrow}, the
problem of intransitivity may pose a problem to choice in democratic
elections. McGarvey\cite{McGarvey1953} proved that any intransitive
preferences on a finite set can be represented as a majority superposition
of a finite number of transitive orders. Intransitivities have become more
philosophically accepted in recent times \cite{Intrans1993} and are now
commonly used in physics \cite{Intrans2011}, biology \cite{BioInt2000} as
well as in social and economic studies \cite{Rank2004,Rank2005,BinaryOp2010}.

Conventional thermodynamics is fundamentally transitive and the
thermodynamics of transitive competition is similar to conventional
thermodynamics in this important respect. We cannot expect competitiveness
to increase indefinitely, as it is likely to have some physical constraints
even if an external source of exergy exempts the system from being isolated
and subject to the immediate constraints of conventional thermodynamics. As
shown in the previous section, if the leading element of the distribution
reaches the point of maximal possible rank in transitive competition, any
further development in the system is terminated. In conventional
thermodynamics, this point is represented by the global equilibrium with
maximal entropy or minimal Gibbs/Helmholtz free energy. The highly
competitive group with maximal ranking would prevent any alternatives from a
successful challenge; the system then stops evolving any further. One may
hope that once a transitive equilibrium is reached, the creative hand of
nature changes external conditions in a ``right'' direction so that the
complex development may resume. Unless the complexity of this intervention
is at least comparable with the complexity of the evolving systems, the
long-term efficacy of such intervention seems doubtful. In most cases stable
systems only slightly alter their states to attain a new equilibrium and
compensate for the environmental disturbance. It seems that transitive
description is an oversimplification of the complex (and often cyclic)
behaviours observed in realistic competitive systems. Transitivity of
competitive thermodynamics is not guaranteed a priori and depends on
transitivity of the competition rules. Multiplicity of competitiveness
criteria combined with a rather limited number of outcomes (i.e. winner or
loser) is most likely to produce intransitivities due to the same reasons
that were first discovered in the Condorcet paradox. Systems with
intransitive competition rules must have an external source of exergy or
negentropy \cite{Life1944}, since isolated systems are subject to the
constraints of conventional thermodynamics and must be transitive. Complex
behaviour is known to occur far from equilibrium of conventional
thermodynamics \cite{Prig1977}, since Onsager's reciprocal relations do not
allow for cycles and enforce transitivity close to the equilibrium.

\subsection{Types of intransitivity}

Any intransitive relation has its transitive closure --- another relation
that is transitive and is, as much as possible, close to the original
relation (see Appendix \ref{sA1} for details). We denote the transitive
closure of our original competition rules by ``$\prec _{t}$'', ``$\succ _{t}$%
'' and\ ``$\simeq _{t}$''. It is useful to distinguish the following
possibilities.

\begin{enumerate}
\item  By transitive closure:

\begin{enumerate}
\item  \textbf{Completely intransitive competition}: all elements are
transitively equivalent, i.e. $\mathbf{y}_{p}\simeq _{t}\mathbf{y}_{q}$ for
any $p$ and $q.$ In terms of the original relation, any two elements $p$ and 
$q$ are a part of at least one intransitive loop 
\begin{equation}
\mathbf{y}_{p}\preccurlyeq \mathbf{y}_{1}\preccurlyeq \mathbf{y}%
_{2}...\preccurlyeq \mathbf{y}_{q}\preccurlyeq \mathbf{y}_{1}^{\prime
}\preccurlyeq \mathbf{y}_{2}^{\prime }...\preccurlyeq \mathbf{y}_{p}
\label{comint}
\end{equation}

\item  \textbf{Intransitive competition having a transitive component: }the
transitive closure involves several transitively unequal classes. In this
case, $\mathbf{y}_{p}\succ _{t}\mathbf{y}_{q}$ requires $\mathbf{y}_{p}\succ 
\mathbf{y}_{q}.$
\end{enumerate}

\item  By localisation:

\begin{enumerate}
\item  \textbf{Locally intransitive competition:} intransitive triplets can
be found in the vicinity of any point

\item  \textbf{Intransitive competition with local transitivity}: the domain
of properties can be divided into subdomains so that the competition is
transitive within each subdomain (but not in the whole domain)
\end{enumerate}
\end{enumerate}

If competition has a transitive component, an absolute ranking $r_{t}$
corresponding to this component can be introduced. This ranking, however,
would be the same for all elements from the same transitive class, i.e. $%
r_{t}(\mathbf{y}_{p})=r_{t}(\mathbf{y}_{q})$ for $\mathbf{y}_{p}\simeq _{t}%
\mathbf{y}_{q}$ while it might be the case that $\mathbf{y}_{p}\succ \mathbf{%
y}_{q}$ or $\mathbf{y}_{p}\prec \mathbf{y}_{q}.$ If competition is locally
transitive, rankings can be introduced for each transitive subdomain but can
not be consistently extended to the whole domain. In general, a complex
competition may involve a sophisticated hierarchy of transitive and
intransitive rules. For example, intransitive competition may have a locally
transitive component or intransitive competition may be locally transitive
and have a transitive component, etc.

\begin{figure}[ht] 
\caption{Competitive version of the Condorcet paradox: an example of intransitivity of group co-ranking 
  $[f_{\func{A}}]\prec \lbrack f_{\func{B}}]\prec \lbrack f_{\func{C}}]\prec \lbrack f_{\func{A}}]$ 
  occuring when the underlying competition 
  is strictly transitive and determined by absolute ranking $r_\#$. 
  Competition is presumed to be non-preferential. The square demonstrates that 
  $\lbrack f_{\func{B}}]\prec \lbrack f_{\func{C}}]$ 
  since B wins over C only in 4/9 of all cases. } 
\begin{center} 
\includegraphics[width=9cm]{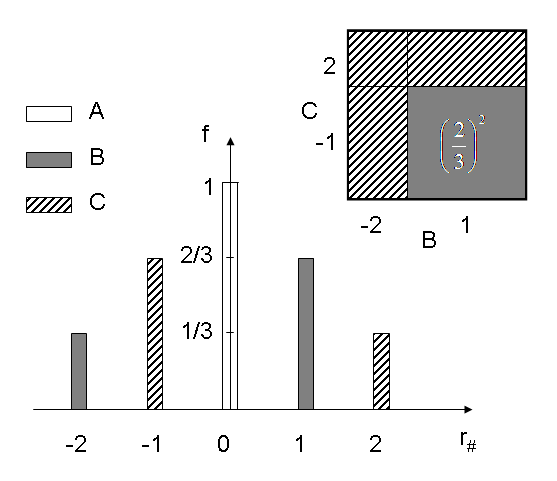}  
\label{fig_A1}  
\end{center}
\end{figure}

Intransitivities may be also distinguished by their origin. A common source
of intransitivity is superimposition of perfectly transitive rules. For
example, comparison of subsystems by co-ranking (that is $[f_{\func{A}%
}]\preccurlyeq \lbrack f_{\func{B}}]$ when $\bar{R}([f_{\func{A}}],[f_{\func{%
B}}])\leq 0$) can exhibit intransitive properties even if the underlined
competition between particles is strictly transitive. This can be
interpreted as a variation of the Condorcet paradox\cite{Cond1785}. The
example of three distributions of particles $f_{\func{A}}$, $f_{\func{B}}$
and $f_{\func{C}}$ such that $[f_{\func{A}}]\prec \lbrack f_{\func{B}}]\prec
\lbrack f_{\func{C}}]\prec \lbrack f_{\func{A}}]$ is shown in Figure \ref
{fig_A1}. We may interpret the sybsystems as competing super-elements but
should expect that the rules for this competition are intransitive
irrespective of the transitivity of the original competition rules.

\subsection{Gibbs mutations in intransitive systems.}

If mutations are restricted to Gibbs mutations, only one equilibrium
distribution $f_{0}(\mathbf{y})$ is possible in the case of complete
intransitivity since mutations defined by equation (\ref{mut-tc}) propagate
to the whole of the domain through the chain (\ref{comint}). Different
equilibriums, however, are possible when the competition has a transitive
component. Since the component ordering denoted by $\succ _{t}$ is
transitive, an absolute ranking can be introduced so that $r_{t}(\mathbf{y}%
_{p})\leq r_{t}(\mathbf{y}_{q})$ is equivalent to $\mathbf{y}%
_{p}\preccurlyeq _{t}\mathbf{y}_{q}$. If intransitive competition has a
transitive component, the system behaves with respect to this component as
discussed in the previous section. The equilibrium distributions can be
written as $f_{0}(\mathbf{y},r_{\ast })$ where $r_{\ast }$ is the ranking of
the leading class and $r_{\ast }$ may increase if some positive mutations
are present in the system. The H-theorem (Theorem \ref{thH1}) apply to Gibbs
mutations irrespective of the transitivity of the competition.

\subsection{Current or local transitivity}

Distribution of a finite number of particles may be confined to a much
smaller region (we can call it current region) as compared to the region of
strict positiveness $f>0$ of the function $f(\mathbf{y})$: particles can not
be found in the region where $f(\mathbf{y})$ is formally positive but very
small. Competition in the current region may be transitive while remaining
intransitive in larger regions. We can characterise this situation as 
\textit{currently transitive} distribution. Currently transitive
distributions behave over short period of time as if the competition is
transitive. This case is illustrated in Figure \ref{fig_4}. The competition
shown in this figure is completely intransitive and an equilibrium
distribution spreads over the whole domain. If the system is invariant with
respect to shifts along the circle, this equilibrium distribution must be
uniform. There is, however, another possibility when the number of particles
is limited: the equilibrium distribution of particles can be confined to a
narrow segment of the circle since $f$ in the rest of the domain is too
small to be taken into account. If some rare positive mutations are present
in the system, this distribution will keep cycling around the circle
indefinitely. In this example, the competition is completely intransitive
but currently transitive.

\subsection{General mutations in intransitive systems.}

\begin{figure}[ht] 
\caption{Current transitivity in completely intransitive competition.} 
\begin{center} 
\includegraphics[width=9cm]{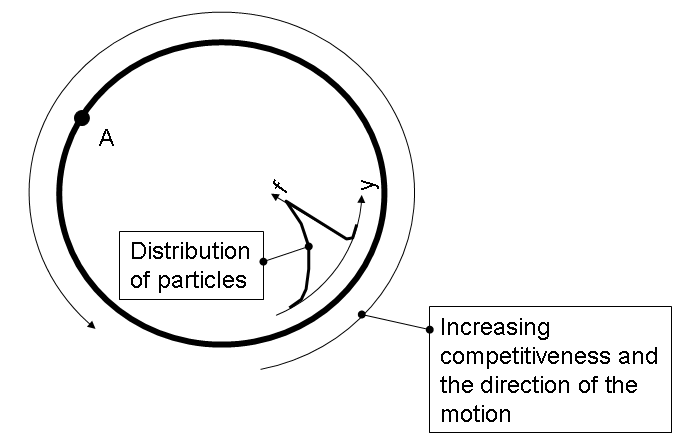}  
\label{fig_4}  
\end{center}
\end{figure}

In transitive competition, systems with general mutations behave in a
qualitatively similar way as compared to systems with Gibbs mutations. This
is not necessarily the case when competition is intransitive. In the absence
of mutations, all non-trivial stationary distributions tend to produce
oscillations (unless $R_{pq}=0$ for all non-isolated $p$ and $q$ --- see
Appendix \ref{sA5}). Instabilities can also be expected in intransitive
systems if the level of mutations is insufficient. Although there is a large
diversity of possibilities in intransitive competitions, the behaviour of
the systems may be predicted when certain restrictions apply. Intransitivity
may be weak and dominated by transitive relations. If competition is
intransitive but has a transitive component, the system would behave with
respect to this component in a way that is similar to the case of general
mutations in transitive systems. The ranking associated with the transitive
component can stay constant or increase in time. A similar behaviour can be
expected for competition that is currently transitive. The distribution
evolves as if the competition is transitive over a short period of time but
the system may appear to be cyclic over longer periods as illustrated in
Figure \ref{fig_4}. Absolute ranking can be introduced within a sector of
the circles in Figure \ref{fig_4} but not over the whole domain. Cyclic
behaviour is common for intransitive competition: at any given moment the
system seems to progress forward but after the cycle is completed, it finds
itself in the original state. Changes that seem to be improvements at a
given time may prove to be detrimental in the long run.

Although intransitive competition does not guarantee a global improvement in
competitiveness of a system, we still may expect some degree of local
consistency with competitive thermodynamics. Since there is generally no
absolute ranking in intransitive competition, we can use a relative ranking
measured with respect to current distribution $f^{\circ }=f(\mathbf{y}%
,t^{\circ })$ that is $r^{\circ }(\mathbf{y)}=r(\mathbf{y},[f^{\circ }])$
and $\bar{R}^{\circ }([f])=\bar{R}([f],[f^{\circ }])$. The distribution $%
f^{\circ }$ remains fixed and, as $f(\mathbf{y},t)$ evolves, $f$ becomes
more and more different from $f^{\circ }$. We, however, consider a
short-term development of $f,$ when $f$ stays close to $f^{\circ }.$
According to the evolution equation (\ref{c-e}), the competition step $%
\delta f_{c}$ always improves current ranking $\bar{R}^{\circ }([\delta
f_{c}])>0,$ but tends to decrease the configurational entropy $S_{c}$. The
following mutation step $\delta f_{m}$ tend to decrease ranking $\bar{R}%
^{\circ }([\delta f_{m}])<0,$\thinspace since most of the mutations are
non-positive and increase the configurational entropy $S_{c}$, since
mutations are random. In a steady case $f=f_{0}$ all these changes
compensate each other and entropy defined by equation (\ref{Sf}) reaches
its\ maximum. It is possible, however, to have a nearly steady state which
is stable but continues to evolve slowly, that is $f_{0}=f_{0}(\mathbf{y,y}%
_{\ast }^{\circ })$ where $\mathbf{y}_{\ast }^{\circ }$ is the $f^{\circ }$%
-graded leading particle (the particle from the distribution $f$ with the
maximal ranking $r_{\ast }^{\circ }=r(\mathbf{y}_{\ast }^{\circ },[f^{\circ
}])$ ). In many cases $r_{\ast }^{\circ }=r_{\ast }^{\circ }(t)$\ tends to
increase in time and conditions sufficient to ensure this can be nominated
(for example possessing a current transitivity or a transitive component).
It seems, however, that cases of decreasing $r_{\ast }^{\circ }(t)$ are also
possible in some circumstances: we call these cases \textit{competitive
degradations}. There are indications (Section \ref{sec_EX4}) that
degradations are accompanied by an increase in chaotic behaviour and it may
be the case that properly defined overall entropy still increases, although
there is no certainty. Competitive degradations, which are competition
failures, should be distinguished from erosive degradations, which are
induced by physical inabilities of the system to retain information as
needed or by partial suppression of the competition.

The intransitive competition process is blind and can not guarantee long
term absolute increase in rankings. Ranking in intransitive competition
becomes relative: what seems to be a competitive improvement now may later
appear to be a loss of competitiveness. It is natural, however, that
competition improves the current relative ranking $\bar{R}^{\circ }([f])=%
\bar{R}([f],[f^{\circ }])$. The example shown in Figure \ref{fig_4}
illustrates the case when the current ranking $\bar{R}^{\circ }([f])$
increases as the distribution $f(\mathbf{y,y}_{\ast }^{\circ }(t))$ moves
counterclockwise. This distribution, however, may lose its stability and
collapse due to disturbances located at point A, since ranking of the
distribution with respect to point A decreases as $\mathbf{y}_{\ast }^{\circ
}(t)$ gets closer to A. Hence, decline and collapse may be caused by both of
the factors mentioned above: direct competitive degradations accompanied by
reduction of $\bar{R}^{\circ }([f])$ and by short-term escalation of ranking 
$\bar{R}^{\circ }([f])$ that appears to be detrimental to the stability of
the distribution over a long run.

\section{Examples of different behaviours observed in competitive systems 
\label{sec_EX0}}

Without implying that competitive behaviour must be limited to the modes
listed below, we distinguish the following types of behaviour in realistic
systems:

\begin{enumerate}
\item  \textbf{Stable equilibrium. }A competitive system remains in this
state indefinitely unless surrounding conditions are changed.

\item  \textbf{Competitive escalation.} A competitive system reaches a
quasi-equilibrium state and continues to slowly evolve in the direction of
increasing competitiveness

\item  \textbf{Invasion wave }is a rapid escalation occurring in a
propagating wave. This process is inhomogeneous.

\item  $^{\ast }$\textbf{Regular cycle}. Competitive escalation with respect
to the current distribution results in recurrence of the same conditions due
to cyclic intransitivity.

\item  $^{\ast }$\textbf{Competitive cooperation and self-organisation}.
Under certain conditions when competition is intransitive and localised in
physical space, the elements tend to form cooperative structures (with a
reduced level of competition within the structure) and collectively struggle
for domination over other structures.

\item  $^{\ast }$\textbf{Competitive degradation}. A system reaches a
quasi-equilibrium state which slowly loses its competitiveness.

\item  $^{\ast }$\textbf{Leaping cycle}. A structure quickly rises to
dominance, holds a dominant position and then weakens due to competitive
degradation or loss of stability and rapidly collapses. If the competition
has a transitive component, each subsequent cycle should be somewhat
different from the previous one, so that the evolution of the system
resembles a spiral.

\item  $^{\ast }$\textbf{Unstable and/or} \textbf{stochastic behaviour}. A
system does not have a stable equilibrium (or quasi-equilibrium) and
possibly evolves in an irregular manner.
\end{enumerate}

The types of behaviour marked by the asterisk can be observed only in
intransitive systems.\ An example of a stable equilibrium is given by
non-negative mutations in transitive competition. Competitive escalation is
related to positive mutations, whose effect is shown in Figure \ref{fig_3}.
Turbulent premixed flames give an example of invasion waves \cite{KP2012}.
Competitive cooperation, self-organisation and competitive degradation have
been detected to occur in intransitive systems with localisation of mixing
in physical space \cite{K-PS2010}. While transitive behaviour is more
consistent with conventional thermodynamics, which is constrained by its
zeroth law and is fundamentally transitive, complex behaviour in competitive
systems is associated with intransitivity. Examples of intransitive
competitions are considered below.

\subsection{Turbulent flows}

\begin{figure}[ht] 
\caption{Intransitivity of energy exchange between Reynolds stresses in turbulent shear flows} 
\begin{center} 
\includegraphics[width=12cm]{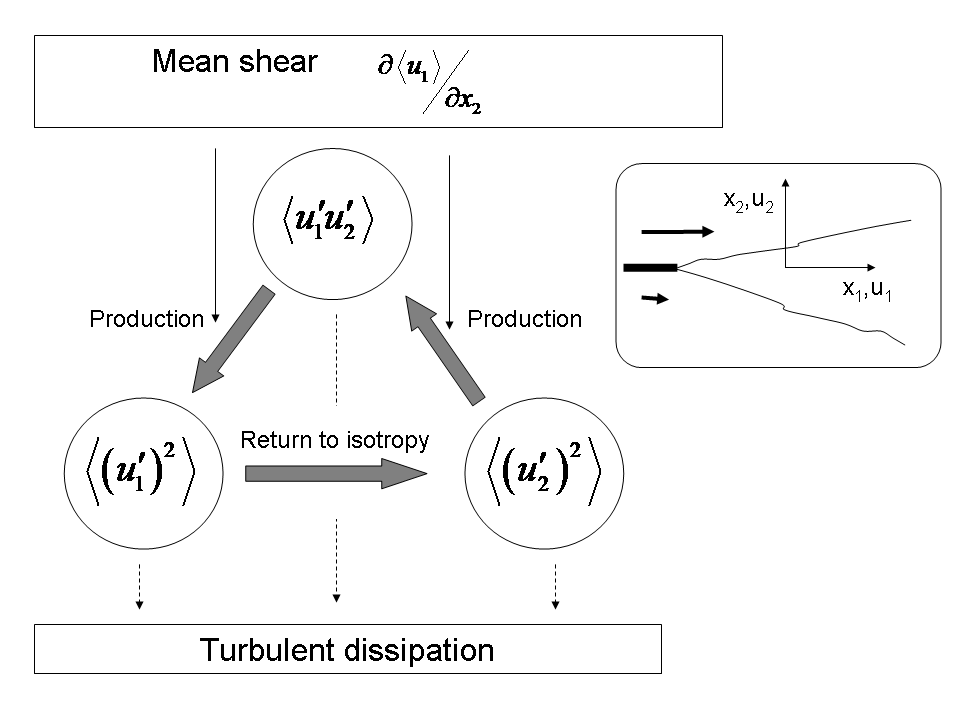}  
\label{fig_5}  
\end{center}
\end{figure}

Turbulence is a complex phenomenon which can provide examples of
intransitive behaviour\footnote{As demonstrated by  van Heijst et. al. \cite{Heijst2007}, 
two-dimensional turbulence 
can display complex self-organising (and likely intransitive) behaviour, 
which appears to be very similar 
to behaviour described for complex competitive systems in Section 7.4}. 
One of the possible examples of turbulent
intransitivities is illustrated in Figure \ref{fig_5}, which shows the
routes for kinetic energy exchange between different Reynolds stresses in a
turbulent shear flow\cite{Pope-00}. Energy is supplied from the mean shear
and dissipated at the smallest scales. This supply/dissipation process,
however, is not uniform: competition for energy between different Reynolds
stresses occurs in an intransitive manner as shown in Figure \ref{fig_5}. In
general, the cyclic nature of the energy exchange shown in the figure may
produce oscillations during rapid distortions but it seems that dissipation,
which is strong in turbulent flows, tends to dampen these oscillations in
most cases.

\subsection{Belousov-Zhabotinsky reaction}

\begin{figure}[ht] 
\caption{Intransitivity of the Oregonator model \cite{Oregonator75} for the
Belousov-Zhabotinsky reaction.} 
\begin{center} 
\includegraphics[width=12cm]{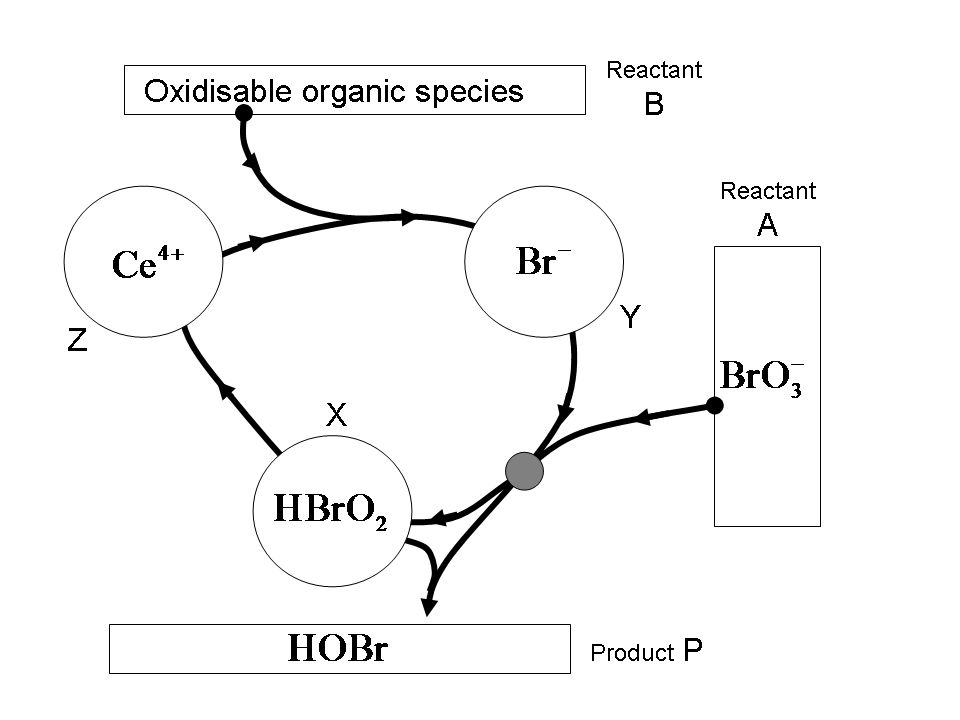}  
\label{fig_6}  
\end{center}
\end{figure}

Intransitivities are possible in chemical reactions. The
Belousov-Zhabotinsky reaction is known to display a cyclic chemical behavior
in a homogeneous mixture, which is unusual since most chemical kinetics tend
to monotonically converge to an equilibrium or steady state. If the
reactants denoted by A and B are supplied to the system and the product P is
removed from the system, the oscillations of concentrations can continue
indefinitely. The simplest chemical kinetic scheme that can adequately
simulate the Belousov-Zhabotinsky reaction is Oregonator \cite{Oregonator75}%
. This scheme involves an essential intransitivity related to the following
cycle X$\rightarrow $Z$\rightarrow $Y$\rightarrow $X$\rightarrow $...,
schematically depicted in Figure \ref{fig_6}. It should be noted that this
cycle has to be powered by external sources of exergy (supply of reactant A
and B in this case) since autonomous conversion of X into Z then into Y then
into X is impossible due to transitive constraints of conventional
thermodynamics.

Although the Oregonator system involves only three active elements X, Y and
Z, its cycle deviates from a regular cycle and has some features of the
leaping cycle: the phase of rapid expansion X$\rightarrow $Z$\rightarrow $Y
is followed by a slow decay of Y until the system loses stability and a new
pulse of injection of X repeats the cycle.

\subsection{Oscillations in the scissors--paper--rock system}

\begin{figure}[ht] 
\caption{Regular cycle in intransitive competition without mutations}
\begin{center} 
\includegraphics[width=7cm]{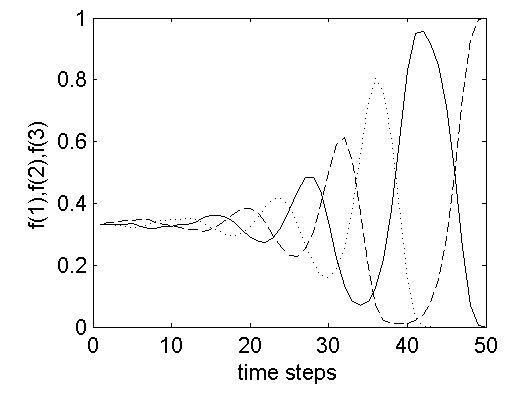}  
\label{fig_6a}  
\end{center}
\end{figure}

Consider a system that has three states: 1) paper, 2) scissors\ and\ 3)
rock. The competition rules are intransitive and given by $1\prec 2\prec
3\prec 1$. No mutations are present in the system. The initial distribution
is given by $f(1)=f(2)=f(3)=1/3$. This distribution is stationary due to the
symmetry of the competition rules. According to Appendix \ref{sA5},
oscillations are expected in this system, while finite time steps make these
oscillations mildly unstable. Figure \ref{fig_6a} demonstrates evolution of
this system.

\subsection{Intransitive competition and competitive cooperation \label%
{sec_EX4}}

The figures presented below are obtained from computer simulations of
abstract competition. In these simulations,\ particle properties are
represented by the vector $\mathbf{y}=(y^{(1)},$ $y^{(2)},$ $y^{(3)})$,
which satisfies the conservation constraint $y^{(1)}+y^{(2)}+y^{(3)}=1$ and
is interpreted as a combination of the primary colors: red, green and blue.
The competition is controlled by the Condorcet rules representing the
majority superposition of the following orderings: $y_{p}^{(i)}<y_{q}^{(i)}$%
, $i=1,2,3$ for every given couple $p$ and $q$. These rules are locally
intransitive and intransitive triplets $\mathbf{y}_{p}\mathbf{\prec y}_{q}%
\mathbf{\prec y}_{r}\mathbf{\prec y}_{p}$ can be found in a vicinity of any
point $p$ in the property space. The competition process is localised in the
physical space. The physical domain is 2-dimensional and is mapped into a
rectangle in the images presented here. None of the primary colors has any
competitive advantages over the other primary colors but particles may have
different overall competition strength with brighter colors performing on
average better than the darker colors. Particle colors change according to
the competition rules and predominantly negative random mutations.
Otherwise, there is no coordination of any kind over the particle
properties. The competition rules and process parameters do not change
during the simulations. The details of the methodology of these competitive
simulations can be found in Ref.\cite{K-PS2010}. We stress that the complex
behavior observed in these simulations is linked to intransitivity and
localisation of the competition.

The formation of a spot-like structure, which is observed in intransitive
simulations with localisation of competition effective volume in physical
space, reduces the intensity of competition. This reduction can be
interpreted as a \textit{competitive cooperation} between the particles. The
particles of a dominant color manage to dominate collectively while reducing
their overall competition effort. The intensity of competition is the
average magnitude of adjustments of the properties of the losers $\Xi
=2\Sigma _{p}\left| \mathbf{\acute{y}}_{p}-\mathbf{y}_{p}\right| /n,$ where
the sum is taken over $n/2$ losing particles. There are two major factors
that reduce the competition intensity. The competition is, obviously, less
intense within a spot filled with similar colors than between the spots
having very different colors. Within a spot, particles of different
competitive strengths tend to compete at their own level of strength.

\begin{figure}[ht] 
\caption{Average intensity of colours indicating leaping cycles, average relative rank $\bar{R}_u$, configurational entropy $S_c$ and average intensity of competition $\Xi$ 
indicating the level of competitive cooperation in intransitive abstract competition. 
The vertical dashed line shows location of images displayed in the next Figure.} 
\begin{center} 
\includegraphics[width=9cm]{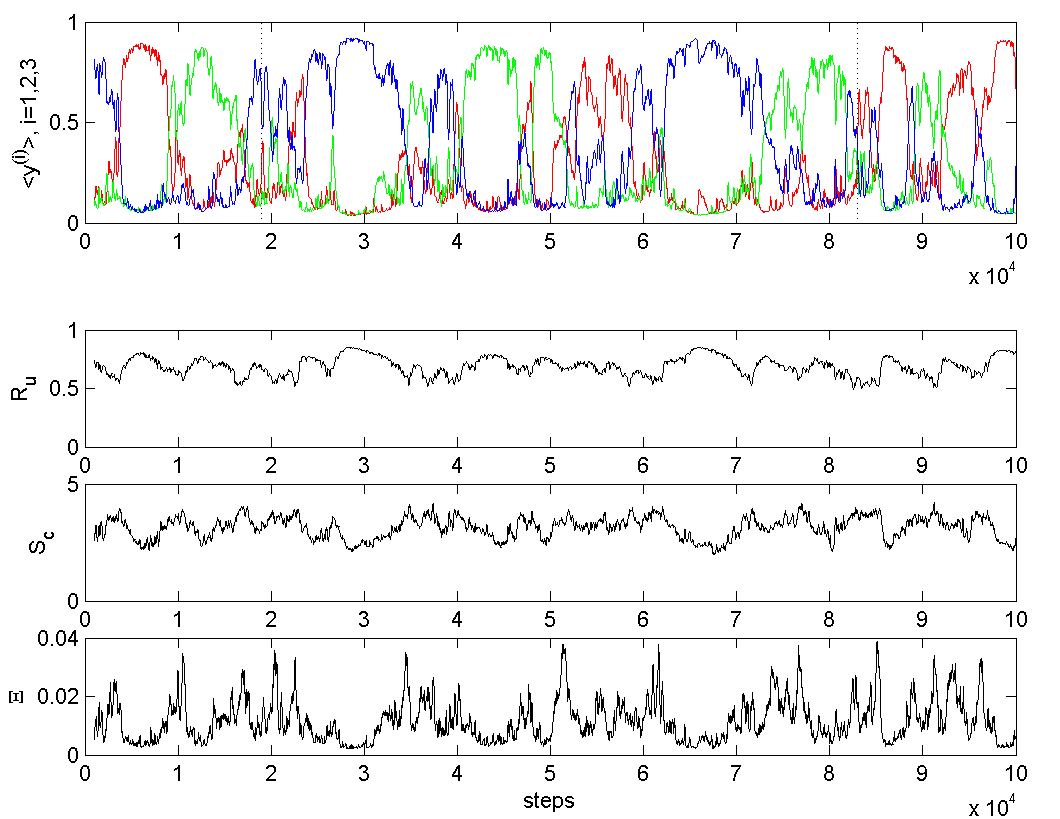}  
\label{fig_7}  
\end{center}
\end{figure}

\begin{figure}[ht] 
\caption{Order appearing out of chaos: the red spot gradually asserts its dominance over the domain that previously was in a state of chaos. 
  The blue spot on the left attempts to launch a competitive bid for dominance but ultimately fails. Two consecutive frames of this process are shown. 
  The property space is shown in the left column of figures while corresponding physical spaces are shown in the right column.} 
\begin{center} 
\includegraphics[width=10cm]{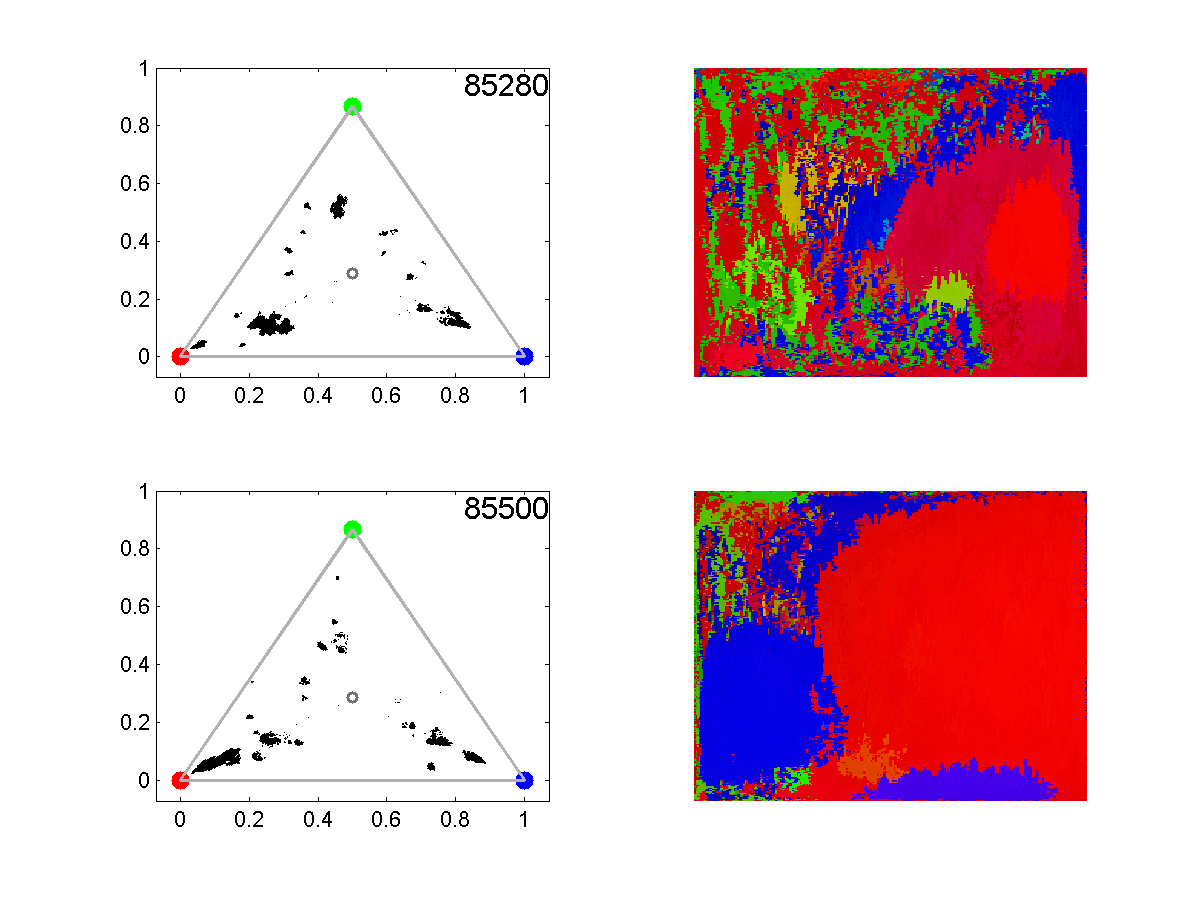}  
\label{fig_8}  
\end{center}
\end{figure}

The curves in Figure \ref{fig_7} represent the intensity of the primary
colors $y^{(i)}$, $i=1$ (red), 2 (green) and 3 (blue) averaged over all
particles and plotted versus the number of time steps. The simulations
display \textit{leaping cycles}: a color leaps into a dominant position of
controlling the domain and manages to fend off competitors for a while, but
ultimately reduces its competitiveness due to \textit{competitive degradation%
} and loses competition to new challenges. The periods of clear dominance of
a color are interchanged with periods of chaotic struggle without a clear
winner or by periods of dominance of another color. The figure also presents
configurations entropy $S_{c}$ determined for $A=1,$ ranking $\bar{R}%
_{u}([f])=\bar{R}([f],[f_{u}])$ relative to the uniform distribution $f_{u}=%
\func{const}$ and the intensity of competition $\Xi $. It is clear that the
loss of ranking $\bar{R}_{u}$ brings a more chaotic behaviour into the
system -- negative correlation between $S_{c}$ and \ $\bar{R}_{u}$ is
prominent. More chaotic behaviour tends to increase the intensity of
competition $\Xi $. Figure \ref{fig_8} shows two consecutive states of the
system (the property space on the left and the physical space on the right)
when a newborn structure takes control over a chaotic behaviour and becomes
highly dominant until it gradually weakens and loses its dominant position.
Note the pyramid-like cooperative structures in the property space ---
within these structures the competition tends to be stratified and occur
between particles with close properties and ranks.

\section{Summary and conclusions}

This work reviews existing publications and introduces a number of new
results related to the application of non-conventional thermodynamics to
turbulence, combustion and general non-equilibrium competitive processes
involving different forms of mixing.

\subsection{Competitive mixing.}

The possibility of application of general principles of thermodynamics to
competitive systems is reviewed and investigated. In its abstract form,
competition is essentially a type of mixing, (i.e. competitive mixing),
which can be effectively represented by a system of Pope particles and used
to simulate a wide range of non-equilibrium processes including turbulent
mixing and combustion. Although abstract competition should have a wide
interdisciplinary range of applications, examples mentioned in this work are
mainly restricted to turbulent flows, chemical reactions and combustion,
which are the author's prime areas of interest. In applications not related
to turbulence, the concept of competitiveness represents a logical extension
of the concepts of fitness and/or utility. Although abstract competition has
been derived from a single field (turbulent combustion), it seems to
represent an interdisciplinary approach linking distant fields and concepts
(adiabatic accessibility and economic preferences may serve as an example).

\subsection{Competitive thermodynamics.}

Thermodynamic description of stochastic systems can be very useful and is
repeatedly used in the literature but, in application to complex systems,
thermodynamics does have its limitations discussed below. Entropy is the key
quantity in thermodynamics, while the other thermodynamic properties are
linked to the definition of entropy. During last decades it has become
conventional to use the term entropy in different contexts, ranging from
mathematics to social sciences. The same trend can be observed in studies of
turbulent flows, where different physical quantities may be implied under
this term. In the present work we use the term ``apparent'' to distinguish
entropy-like quantities from molecular entropy.

We introduce a special type of random disturbances present in competition
--- the Gibbs mutations which possess certain Markov-like properties --- and
demonstrate that general principles of competition are consistent with
thermodynamic description. Competitive entropy is subject to the competitive
H-theorem presented in Appendix \ref{sA5}. A competitive system can also be
characterised by another thermodynamic quantity --- the competitive
potential, which determines the likely direction of evolution of the system.
A competitive system can be generally expected to stay in equilibrium or
gradually increase its competitive potential over time. Contested resources
tend to move between systems from lower to higher values of the competitive
potential. In premixed combustion, products have higher entropy than the
reactants and thus are more competitive. In the same manner, more
competitive species invading areas occupied by less competitive species
should be assigned a higher value of the effective entropy. Competitive
thermodynamics recognises the obvious trend of moving towards more
competitive states, irrespective of the exact nature of this
competitiveness. The adiabatic accessibility approach, the theory of Gibbs
measures and the principles of non-equilibrium thermodynamics should be
mentioned here as important results preceding competitive thermodynamics.
The overall rate of selection is linked to the frequency of positive
mutations, which is constrained by the fluctuation theorem and other
thermodynamic parameters.

The analogy with conventional thermodynamics is established first for Gibbs
mutations but, if competition is transitive, any system with weakly positive
mutations of a general nature would behave in a qualitatively similar manner
to Gibbs systems: the system would quickly reach its quasi-equilibrium and
then slowly escalate towards more competitive states until the point of
maximal competitiveness is reached. After this, no further evolution can
occur in the system as long as the external conditions remain fixed. In
principle, a transitive competitive system, which is decoupled from
molecular thermodynamics by an external source of exergy, may escalate its
competitiveness indefinitely but, practically, there should be some physical
limits on how high this competitiveness could be. Any equilibrium represents
a balance between the forces of chaos and the forces of order. Thus, from
the perspective of competitive thermodynamics, some degree of order
appearing out of chaos in competitive systems should be expected in the same
way as more ordered crystalline states can appear in conventional
thermodynamics under appropriate conditions. From the perspective of
competitive thermodynamics, order appearing in some conventionally
non-equilibrium competitive processes is, by itself, no more or less
surprising than the possible presence of order in equilibrium states of
conventional thermodynamics.

\subsection{Intransitivity.}

The main argument presented in this paper is, however, not about
similarities of conventional and competitive thermodynamics, but about their
differences. In the case of relatively simple systems, transitive
competition and transitive thermodynamics produce a plausible picture of
monotonic motion towards competitive equilibrium. This, however, does not
seem to be consistent with the endlessly vigorous and often cyclic behaviour
observed in complex competitive systems. However, if the competition rules
are allowed to be intransitive, new types of complex behaviours emerge
within the systems. The elements may form structures with a reduced level of
competition within the structure (i.e. competitive cooperation) and struggle
collectively to dominate the allowable space. Endless cyclic behaviour seems
to become quite common under intransitive conditions: cooperative structures
tend to survive and dominate for some time only to degrade at the end and be
replaced by new structures. Intransitive systems can not be
thermodynamically isolated, as this would immediately impose the transitive
constrains of conventional thermodynamics, but require a relatively simple
form of intervention --- an external source of exergy. The choice in favour
of considering intransitivities needs to be made not because intransitivity
can immediately explain all the complexities of the surrounding world but
because transitive competition certainly cannot. This does not prevent many
specific features from having perfectly transitive explanations, while
intransitivity is associated with more complex effects.

The early success of classical thermodynamics was associated with
recognition of the irreversibility of the world around us, in spite of the
time reversibility adopted by classical physics. It is difficult to make a
general statement but abstract competition seems to point towards
intransitivity, and the number of publications dedicated to different
aspects of intransitivity is on the rise. We thus might have to face another
difficult task of recognising the intransitivity of the surrounding world in
contrast with the natural preference for the simplicity of the transitive
description. In the transitive world, we have clear signposts: ``strong''
versus ``weak'', ``fit'' versus ``unfit'' and so on. The intransitive word
is much more relativistic: a strategy that seems to be a big winner today
may prove to be disastrous in a long run. We do not like intransitivities,
not because they do not exist, but because they cause complications and it
is generally difficult to treat them in a logical manner. This, however, is
exactly the reason why intransitivities should be studied --- they are
responsible for the complexity of the world around us. The reasons behind
the ubiquitousness of intransitivity seem to be quite transparent and very
much similar to the reasons behind the Condorcet paradox: intransitivity is
common when the simple outcome ``is'' or ``is not'' depends on many factors
and criteria. A complex system is characterised by a multiplicity of rules;
even if each rule is perfectly transitive, it is quite likely that a
superposition of these rules is not. An element of a system may represent a
system on its own (i.e. subsystem) while co-ranking of subsystems is
inherently intransitive (see the example shown in Figure \ref{fig_A1}).
Another cause of intransitivity is that transitively weak elements tend to
be removed by competition and further competition occurs in a more refined
domain, where elements have similar ranks and intransitivities are likely.

\subsection{Unresolved problems.}

The theorems given in the Appendices establish major principles for
evolution of competitive systems. It is worthwhile, however, to draw the
attention of the reader to the immediate unresolved problems associated with
intransitive behaviour. While competitive thermodynamics is useful when
competition is intransitive but retains some transitive properties (such as
being currently transitive or having a transitive component) or when
mutations are not much different from Gibbs mutations, the feasibility of
thermodynamic treatment of general mutations combined with general
intransitivity is under the question mark. The analogy with conventional
thermodynamics works well for simpler competitions but it\ seems to become
less suitable for complex cases.

Competitive degradations are an outcome of competition that seems abnormal
from the perspective of apparent thermodynamics, but they appear to be
possible in intransitive competitions. It seems that competitive
cooperation, which also becomes possible in intransitive systems and is one
of the key signs of complex behaviour, has a penalty of competitive
degradation attached to the benefits of competitive cooperation.
Establishing necessary and sufficient conditions for the existence of
competitive degradations is an outstanding problem.

\subsection{Conclusion}

Competitive thermodynamics, whose concept has been derived from the
modelling of turbulent combustion by competitive mixing, indicates that
transitive evolution of competitive systems is generally consistent with
conventional thermodynamic principles. Studies of complex behaviour in
competitive systems, however, need to move beyond the principles of
conventional thermodynamics in order to take into account intransitivity,
which is commonly present in nature and seems to be responsible for complex
cooperative behaviour.

\section{Acknowledgments}

The author thanks D.N.P. Murthy for numerous discussions and S.B.Pope for
insightful comments and suggestions. Suggestions made by the anonymous
reviewers are also appreciated. The author thanks D.N. Saulov and D.A.
Klimenko for useful comments and assistance in preparation of the
manuscript. The wide scope of possible applications of this work is likely
to make some omissions inevitable; any omissions, irrespective of their
fields of application, are regretted by the author. The part of this work
related to particle methods is supported by the Australian Research Council.


\begin{appendices}



\begin{center}
{\LARGE APPENDICES}
\end{center}


\section{Transitivity and absolute ranking\label{sA1}}

This section gives a brief explanation of the mathematical terms and
statements referred to in the main text.

\subsection{Absolute ranking and Debreu theorem}

Consider totally and transitively pre-ordered sets in Euclidean space or in
a complete separable metric space:

\begin{itemize}
\item  \textbf{Totality} implies that any two elements $p$ and $q$ are
comparable, that is either $\mathbf{y}_{p}\preccurlyeq \mathbf{y}_{q}$ or $%
\mathbf{y}_{p}\succcurlyeq \mathbf{y}_{q}$ or both of these relations are
correct and $\mathbf{y}_{p}\simeq \mathbf{y}_{q}$.

\item  \textbf{Pre-ordering}, unlike ordering, allows for $\mathbf{y}%
_{p}\simeq \mathbf{y}_{q}$ when $\mathbf{y}_{p}$ and $\mathbf{y}_{q}$ are
not the same.

\item  \textbf{Transitivity} is defined as the following property: $\mathbf{y%
}_{p}\preccurlyeq \mathbf{y}_{q}$ and $\mathbf{y}_{q}\preccurlyeq \mathbf{y}%
_{r}$\ always demands that $\mathbf{y}_{p}\preccurlyeq \mathbf{y}_{r}$\ for
any $p,$ $q$ and $r$.

\item  \textbf{Continuity.} Pre-ordering is continuous when the subsets $%
\frak{D}_{\preccurlyeq }(\mathbf{y}^{\circ })=\{\mathbf{y}\;|\;\mathbf{y}%
\preccurlyeq \mathbf{y}^{\circ }\}$ and $\frak{D}_{\succcurlyeq }(\mathbf{y}%
^{\circ })=\{\mathbf{y}\;|\;\mathbf{y}\succcurlyeq \mathbf{y}^{\circ }\}$
are closed for any $\mathbf{y}^{\circ };$ or equivalently when for any
converging sequence $\mathbf{y}_{p},$ $p=1,2,...$ selected from the set, the
boundedness of the sequence by $\mathbf{y}_{p}\preccurlyeq \mathbf{y}^{\circ
}$ (or $\mathbf{y}_{p}\succcurlyeq \mathbf{y}^{\circ }$) demands the
corresponding boundedness of the limit $\mathbf{y}_{p}\rightarrow \mathbf{y}%
_{0}$ as$\;p\rightarrow \infty $ by $\mathbf{y}_{0}\preccurlyeq \mathbf{y}%
^{\circ }$ (or $\mathbf{y}_{0}\succcurlyeq \mathbf{y}^{\circ }$) for any $%
\mathbf{y}^{\circ }$.\ Physically, continuity of the ordering indicates a
connection between ordering and the intrinsic metric of the space.

\item  \textbf{Absolute ranking}, which is a scalar function $r_{\#}(\mathbf{%
y)}$ defined on the set, is equivalent to pre-ordering when $r_{\#}(\mathbf{y%
}_{p}\mathbf{)\leq }r_{\#}(\mathbf{y}_{q}\mathbf{)}$ is equivalent to $%
\mathbf{y}_{p}\preccurlyeq \mathbf{y}_{q}$ for any $p$ and $q.$ Ranking is
called continuous when the function $r_{\#}(\mathbf{y)}$ is continuous.

\item  \textbf{Debreu theorem.} The introduction of absolute ranking is
subject to the Debreu theorem \cite{Debreu1954,Debreu1964} which states that
equivalent absolute ranking $r_{\#}(\mathbf{y}_{p}\mathbf{)}$ can be
introduced for any transitive total pre-ordering provided the pre-ordering
is continuous. The ranking can be selected to be continuous.
\end{itemize}

Demonstrating the necessity of the continuity restriction imposed on
pre-ordering is relatively easy. If the converging sequence $\mathbf{y}%
_{p}\rightarrow \mathbf{y}_{0}$ as$\;p\rightarrow \infty $ is bounded by $%
\mathbf{y}_{p}\preccurlyeq \mathbf{y}^{\circ }$ then the equivalent
continuous ranks $r_{p}=r_{\#}(\mathbf{y}_{p})$ are bounded by $r_{p}\leq
r^{\circ }=r_{\#}(\mathbf{y}^{\circ })$. The ranks $r_{p}$ converge to $%
r_{p}\rightarrow r_{0}=r_{\#}(\mathbf{y}_{0})$ since the ranking function is
continuous.\ It is clear that $r_{0}\leq r^{\circ },$\ otherwise convergence
of the ranks to $r_{0}$ is impossible. Hence, we conclude that $\mathbf{y}%
_{0}\preccurlyeq \mathbf{y}^{\circ }$. The necessity of transitivity is also
obvious. Indeed, let $\mathbf{y}_{p}\prec \mathbf{y}_{q}\prec \mathbf{y}%
_{r}\prec \mathbf{y}_{p}$ be an intransitive triplet. For equivalent ranking
we then have $r_{\#}(\mathbf{y}_{p})<r_{\#}(\mathbf{y}_{q})<r_{\#}(\mathbf{y}%
_{r})<r_{\#}(\mathbf{y}_{p})$ and ranking in intransitive competition, if it
exists, becomes a multivalued function.

The sufficiency part of the theorem, which is much more difficult to
establish, has been proved by Debreu \cite{Debreu1954,Debreu1964} for
topological second-separable spaces. The results of Lieb and Yngvason \cite
{EntOrd2003}, who also postulated transitivity and continuity (although
called the latter ``stability'') for pre-ordering by adiabatic accessibility
that, are consistent with the Debreu theorem.

The following examples illustrate the possibility or impossibility of
absolute ranking:

\begin{itemize}
\item  \textbf{Lexicographic ordering}. The example of a transitive
preordering that is not continuous and can not have an equivalent ranking
has been given by Debreu \cite{Debreu1954}. Consider lexicographic ordering
of the plane $\mathbf{y}=(y^{(1)},y^{(2)})$. Let $\mathbf{y}_{p}\preccurlyeq 
\mathbf{y}_{q}$ when, by definition, either $y_{p}^{(1)}<y_{q}^{(1)}$ or $%
y_{p}^{(1)}=y_{q}^{(1)}$ and $y_{p}^{(2)}\leq y_{q}^{(2)}$ implying that $%
\mathbf{y}_{p}=\mathbf{y}_{q}$ when $y_{p}^{(1)}=y_{q}^{(1)}$ and $%
y_{p}^{(2)}=y_{q}^{(2)}$. The subsets $\frak{S}(y^{(1)})=\{(y^{(1)},-\infty
<y^{(2)}<+\infty )\}$ should have non-overlapping intervals of ranks $r_{\#}(%
\frak{S}(y_{1}))\cap r_{\#}(\frak{S}(y_{2}))=\emptyset $ when $y_{1}\neq
y_{2}$. Since there is a non-countable number of subsets $\frak{S}(y)$ and
only a countable number of the intervals $r_{\#}(\frak{S}(y_{1}))$,
equivalent ranking is impossible.

\item  \textbf{Countable sets. }Equivalent ranking can always be introduced
for any totally and transitively pre-ordered countable set. Let $\mathbf{y}%
_{1},\mathbf{y}_{2},...$ be the counting sequence that, of course, is
generally not coincident with the ordering sequence. The first element $%
\mathbf{y}_{1}$ can be assigned any rank, say $r_{\#}(\mathbf{y}_{1})=0$.
Assuming that equivalent ranks have been assigned to first $k$ elements from
the counting sequence, we can always select a rank, which is consistent with
the previous $k$ assignments, for the element numbered $k+1$. Repeating this
procedure for the rest of the counting sequence assigns equivalent ranks to
all elements.

\item  \textbf{Measure-based ranking}. If a measure $\mu $ is defined for
the space under consideration, a relatively\ simple absolute ranking can be
introduced for total transitive pre-ordering by 
\begin{equation}
r_{\mu }(\mathbf{y}^{\circ })=\mu \left( \{\mathbf{y\;}|\;\mathbf{y}%
\preccurlyeq \mathbf{y}^{\circ }\}\right)
\end{equation}
From a practical perspective, the measure-based ranking may be sufficiently
accurate but, generally, it does not represent an equivalent ranking and may
coarsen the original pre-ordering:

\begin{enumerate}
\item  if $\mathbf{y}_{p}\simeq \mathbf{y}_{p}$ then $r_{\mu }(\mathbf{y}%
_{p})=r_{\mu }(\mathbf{y}_{p})$

\item  if $r_{\mu }(\mathbf{y}_{p})<r_{\mu }(\mathbf{y}_{q})$ then $\mathbf{y%
}_{p}\prec \mathbf{y}_{q}$

\item  if $r_{\mu }(\mathbf{y}_{p})=r_{\mu }(\mathbf{y}_{q})$ then $\mathbf{y%
}_{p}\prec \mathbf{y}_{q}$ or $\mathbf{y}_{p}\succ \mathbf{y}_{q}$ or $%
\mathbf{y}_{p}\simeq \mathbf{y}_{q}$
\end{enumerate}
\end{itemize}

\subsection{Transitive closure \label{sA3}}

This subsection gives a brief explanation of transitive closures, which are
commonly used in the analysis of intransitive relations; for more details
see other publications, \cite{BinaryOp2010} for example. The transitive
closure defines a new transitive relation ``$\preccurlyeq _{t}$'' that is
consistent as much as possible with a given intransitive relation ``$%
\preccurlyeq $''. Transitive closure is minimal i.e. additional relations,
which are not needed to establish transitivity of existing relations, are
not included in the closure. If ``$\preccurlyeq $'' is transitive than ``$%
\preccurlyeq _{t}$'' coincides with ``$\preccurlyeq $''.

The closure involves several steps.\ For the sake of simplicity, we may
presume that all elements are comparable so that elements in every pair are
related by ``$\prec $'', ``$\succ $'' or ``$\simeq $''. These relations are
first expressed in terms of ``$\preccurlyeq $'', for example ``$\prec $''
means ``$\preccurlyeq $'' but not ``$\succcurlyeq $''. Second, $\mathbf{y}%
_{p}\preccurlyeq _{t}\mathbf{y}_{q}$ is defined for any $p$ and $q$ that $%
\mathbf{y}_{p}\preccurlyeq \mathbf{y}_{q}$. Third, the new relation is
transitively extended to new elements: whenever $\mathbf{y}_{p}\preccurlyeq
_{t}\mathbf{y}_{q}\preccurlyeq _{t}\mathbf{y}_{r}$ it is also set that$\ 
\mathbf{y}_{p}\preccurlyeq _{t}\mathbf{y}_{r}$. The third step is repeated
as long as necessary until these extensions do not introduce any new
relations. Finally, $\mathbf{y}_{p}\simeq _{t}\mathbf{y}_{q}$ is defined
when $\mathbf{y}_{p}\preccurlyeq _{t}\mathbf{y}_{q}$ and $\mathbf{y}%
_{p}\succcurlyeq _{t}\mathbf{y}_{q}$ (that is $\mathbf{y}_{q}\preccurlyeq
_{t}\mathbf{y}_{p}$)$;$ while $\mathbf{y}_{p}\prec _{t}\mathbf{y}_{q}$ is
defined when $\mathbf{y}_{p}\preccurlyeq _{t}\mathbf{y}_{q}$ but not $%
\mathbf{y}_{p}\succcurlyeq _{t}\mathbf{y}_{q}.$ The defined closure is
transitive by definition.

The transitive closure divides all elements into one or more classes of
equivalence $\frak{C}_{I}$ where $\mathbf{y}_{p}\simeq _{t}\mathbf{y}_{q}$
when and only when $p$ and $q$ belong to the same class $I$. The classes are
totally ordered by ``$\prec _{t}$'' so that $\mathbf{y}_{p(I)}\prec _{t}%
\mathbf{y}_{q(J)}$when and only when $\frak{C}_{I}\prec _{t}\frak{C}_{J},$
where $p(I)$ denotes any $p$ that belongs to class $I$. The transitive and
intransitive relations are related to each other by:

\begin{itemize}
\item  Within each class any two elements $p$ and $q$ are part of at least
one loop with a changing but finite number of elements 
\begin{equation}
\mathbf{y}_{p}\preccurlyeq \mathbf{y}_{1}\preccurlyeq \mathbf{y}%
_{2}\preccurlyeq ...\preccurlyeq \mathbf{y}_{q}\preccurlyeq \mathbf{y}%
_{1}^{\prime }\preccurlyeq \mathbf{y}_{2}^{\prime }\preccurlyeq
...\preccurlyeq \mathbf{y}_{p}
\end{equation}
where all elements belong to the same class $\mathbf{y}_{p}\simeq _{t}%
\mathbf{y}_{1}\simeq _{t}...\simeq _{t}\mathbf{y}_{q}\simeq _{t}\mathbf{y}%
_{1}^{\prime }\simeq _{t}...\simeq _{t}\mathbf{y}_{p}$. The shortest
possible loop has only two elements $\mathbf{y}_{p}\simeq \mathbf{y}_{p}$.

\item  The transitive relation coarsens the original relation , that is\ for
any $p$ and $q$

\begin{enumerate}
\item  if $\mathbf{y}_{p}\simeq \mathbf{y}_{p}$ then $\mathbf{y}_{p}\simeq
_{t}\mathbf{y}_{p}$

\item  if $\mathbf{y}_{p}\prec _{t}\mathbf{y}_{q}$ then $\mathbf{y}_{p}\prec 
\mathbf{y}_{q}$

\item  if $\mathbf{y}_{p}\simeq _{t}\mathbf{y}_{q}$ then $\mathbf{y}%
_{p}\prec \mathbf{y}_{q}$ or $\mathbf{y}_{p}\succ \mathbf{y}_{q}$ or $%
\mathbf{y}_{p}\simeq \mathbf{y}_{q}$
\end{enumerate}
\end{itemize}


\section{Relative ranking in preferential competitions \label{sA2}}

This section gives a more general definition of ranking suitable for
preferential mixing. These rankings are weighted by the preference function $%
\Psi _{pq}=\Psi (\mathbf{y}_{p},\mathbf{y}_{q}),$ which determines the
probability of selecting the couple $p,q$ for mixing. This function is
presumed bounded $0\leq \Psi (\mathbf{y}_{p},\mathbf{y}_{q})\leq 1.$ If
mixing is non-preferential $\Psi =1$ for all particles; if particles $p$ and 
$q$ are isolated and can not mix $\mathbf{y}_{p}\parallel \mathbf{y}_{q}$
then $\Psi _{pq}=0$.

\subsection{Losing and winning capacities}

Relative ranking is related to losing and winning capacities. The
generalised Heaviside step function 
\begin{equation}
H(\mathbf{y},\mathbf{y}^{\prime })=\frac{R(\mathbf{y},\mathbf{y}^{\prime })+1%
}{2}=\left\{ 
\begin{array}{c}
0,\;\;\mathbf{y}\prec \mathbf{y}^{\prime } \\ 
\frac{1}{2},\;\;\mathbf{y}\simeq \mathbf{y}^{\prime } \\ 
1,\;\;\mathbf{y}\succ \mathbf{y}^{\prime }
\end{array}
\right.
\end{equation}
has the following properties 
\begin{equation}
H(\mathbf{y},\mathbf{y}^{\prime })=1-H(\mathbf{y}^{\prime },\mathbf{y})
\end{equation}
\begin{equation}
R(\mathbf{y},\mathbf{y}^{\prime })=H(\mathbf{y},\mathbf{y}^{\prime })-H(%
\mathbf{y}^{\prime },\mathbf{y}),
\end{equation}
We denote $H_{pq}=H(\mathbf{y}_{p},\mathbf{y}_{q})$ and define winning and
losing capacities of a particle by relationships 
\begin{equation}
h^{+}(\mathbf{y}_{p}\mathbf{,[}f])=\stackunder{\infty }{\int }\Psi (\mathbf{y%
}_{p},\mathbf{y}_{q})H(\mathbf{y}_{p},\mathbf{y}_{q})f(\mathbf{y}_{q})d%
\mathbf{y}_{q}=\frac{1}{n}\sum_{q}\Psi _{pq}H_{pq}
\end{equation}
\begin{equation}
h^{-}(\mathbf{y}_{p}\mathbf{,[}f])=\stackunder{\infty }{\int }\Psi (\mathbf{y%
}_{p},\mathbf{y}_{q})H(\mathbf{y}_{q},\mathbf{y}_{p})f(\mathbf{y}_{q})d%
\mathbf{y}_{q}=\frac{1}{n}\sum_{q}\Psi _{pq}H_{qp}
\end{equation}
Here, as an example, we give both the continuous and discrete forms,
assuming that the distribution $f(\mathbf{y}_{q})$ is represented by a large
number of particles $q=1,...,n$. The average winning capacity of one
distribution $f_{1}(\mathbf{y})$ with respect to another $f_{2}(\mathbf{y})$ 
\begin{equation}
\bar{H}([f_{1}],[f_{2}])=\stackunder{\infty }{\int \int }\Psi (\mathbf{y}%
_{1},\mathbf{y}_{2})H(\mathbf{y}_{1},\mathbf{y}_{2})f_{1}(\mathbf{y}%
_{1})f_{2}(\mathbf{y}_{2})d\mathbf{y}_{1}d\mathbf{y}_{2}
\end{equation}
represents an average of $h^{+}(\mathbf{y}_{1}\mathbf{,[}f_{2}])$ where $%
\mathbf{y}_{1}$\ is distributed with $f_{1}(\mathbf{y}_{1})$ and $\mathbf{y}%
_{2}$\ is distributed with $f_{2}(\mathbf{y}_{2}),$ while $p_{1}=1,...,n_{1}$
and $p_{2}=1,...,n_{2}$ run over particle groups representing distributions $%
f_{1}$ and $f_{2}$ correspondingly.

\subsection{Relative ranking and co-ranking}

The effective relative ranking of a particle is then given by 
\begin{equation}
r(\mathbf{y}_{p}\mathbf{,[}f])=\frac{h^{+}-h^{-}}{h^{+}+h^{-}}=\frac{\theta (%
\mathbf{y}_{p}\mathbf{,[}f])}{\psi (\mathbf{y}_{p}\mathbf{,[}f])}
\end{equation}
where 
\[
\theta =h^{+}-h^{-}\;\func{and}\;\psi =h^{+}+h^{-} 
\]
is the connectivity of particle $p$ to distribution $f(\mathbf{y})$: that is 
$\psi =1$ if mixing is non-preferential and $\psi =0$ if the particle is
isolated from the distribution $f(\mathbf{y})$ which can be expressed by $%
\mathbf{y}_{p}\parallel \lbrack f]$. Relative $\Psi $-weighted ranking $r$
indicates the strength of particle $\mathbf{y}_{p}$ in competition with
distribution $f(\mathbf{y})$ and $r_{p}=r(\mathbf{y}_{p}\mathbf{,[}f])$ may
not be defined when $\mathbf{y}_{p}\parallel \lbrack f]$.

The relative ranking (co-ranking) of two distributions $f_{1}(\mathbf{y})$
and $f_{2}(\mathbf{y})$ is defined as 
\begin{equation}
\bar{R}([f_{1}],[f_{2}])=\frac{\bar{\Theta}([f_{1}],[f_{2}])}{\bar{\Psi}%
([f_{1}],[f_{2}])}
\end{equation}
where 
\begin{equation}
\bar{\Theta}([f_{1}],[f_{2}])=\bar{H}([f_{1}],[f_{2}])-\bar{H}%
([f_{2}],[f_{1}])
\end{equation}
and 
\begin{equation}
\bar{\Psi}([f_{1}],[f_{2}])=\bar{H}([f_{1}],[f_{2}])+\bar{H}([f_{2}],[f_{1}])
\label{PSIbar}
\end{equation}
is mixing connectivity of these two distributions. If $\bar{\Psi}%
([f_{1}],[f_{2}])=0,$ distributions $f_{1}(\mathbf{y})$ and $f_{2}(\mathbf{y}%
)$ are isolated $[f_{1}]\parallel \lbrack f_{2}]$ and particles from these
distributions do not compete with each other. The equations given above
generalise equations (\ref{rank-1}) -- (\ref{rank-ee}) for preferential
mixing.

\subsection{Subsystem ranking}

If the distribution $f(\mathbf{y})$ is divided into groups or subsystems $%
I=1,...,K$ so that 
\begin{equation}
f(\mathbf{y})=\sum_{I=1}^{K}a_{I}\phi _{I}(\mathbf{y})  \label{grp_f}
\end{equation}
where the distributions $\phi _{I}(\mathbf{y})$ and constants $a_{I}$ are
normalised, i.e. 
\[
\int_{\frak{D}_{I}}\phi _{I}(\mathbf{y})d\mathbf{y}=1,\;\;\;%
\sum_{I=1}^{K}a_{I}=1 
\]
Each group of particles is generally presumed to be a subsystem, which is
confined to a certain domain $\frak{D}_{I}$ distinctive for each subsystem.
The ranking of subsystems can be introduced as 
\begin{equation}
\bar{R}_{I}=\bar{R}([\phi _{I}],[f])=\frac{\bar{\Theta}_{I}}{\bar{\Psi}_{I}}=%
\frac{1}{\bar{\Psi}_{I}}\sum_{J}a_{J}\bar{\Psi}_{IJ}\bar{R}_{IJ}
\end{equation}
\begin{equation}
\bar{\Theta}_{I}=\bar{\Theta}([\phi _{I}],[f])=\sum_{J}a_{J}\bar{\Theta}_{IJ}
\end{equation}
\begin{equation}
\bar{\Psi}_{I}=\bar{\Psi}([\phi _{I}],[f])=\sum_{J}a_{J}\bar{\Psi}_{IJ}
\end{equation}
where $R_{IJ}$ represents the group co-ranking matrix 
\begin{equation}
-\bar{R}_{JI}=\bar{R}_{IJ}=\bar{R}([\phi _{I}],[\phi _{J}])=\frac{\bar{\Theta%
}_{IJ}}{\bar{\Psi}_{IJ}}=\frac{\bar{\Theta}([\phi _{I}],[\phi _{J}])}{\bar{%
\Psi}([\phi _{I}],[\phi _{J}])}
\end{equation}
If the subsystems $I$ and $J$ are isolated $[\phi _{I}]\parallel \lbrack
\phi _{J}]$ then $\bar{\Psi}_{IJ}=0$ and co-ranking of these subsystems $%
R_{JI}$\ is not defined. The possibility of treating the subsystems as new
competing super-elements should be mentioned. Complex systems can be
expected to have a hierarchy of subdivisions.

\section{Mutations \label{sAm}}

This section introduces Gibbs mutations that have some links with to Markov
properties and Gibbs measures. Gibbs mutations are non-negative while
near-Gibbs mutations can be infrequently positive. The distribution of
infrequent positive mutations is analysed using thermodynamic considerations
and the fluctuation theorem.

\subsection{Gibbs mutations}

The case of non-preferential mixing, where connections between Gibbs
mutations, Markovian properties and Gibbs measures are most transparent, is
considered first. A more general definition of Gibbs mutations for
preferential mixing is given at the end of this subsection.

Let $f_{\zeta }(y,y^{\circ }\mathbf{)}$ be the probability distribution of
mutations $y=y^{\circ }+\zeta $ originating at state $y^{\circ }$. Mutations
we consider in this section are non-positive (i.e. $y^{\circ }+\zeta
\preccurlyeq y^{\circ }$) and $f_{\zeta }$ is presumed to be free of
singular components. The function 
\begin{equation}
F_{\zeta }(y,y^{\circ })=\int_{-\infty }^{y}f_{\zeta }(y^{\prime },y^{\circ }%
\mathbf{)}dy^{\prime }  \label{GibbsFm}
\end{equation}
specifies the probability that ranking falls down to $r_{\#}(y)$ or below
for mutations originating at $y^{\circ }$. Note that $F_{\zeta }(y,y^{\circ
})=1$ for $y\succcurlyeq y^{\circ }$ since no positive mutations are
allowed. In addition, the distributions of Gibbs mutations are required to
satisfy the probability decomposition 
\begin{equation}
F_{\zeta }(y,y^{\circ })=F_{\zeta }(y,y^{\prime })F_{\zeta }(y^{\prime
},y^{\circ })  \label{Gibbs3}
\end{equation}
for any $y\preccurlyeq y^{\prime }\preccurlyeq y^{\circ }$. This condition
can be interpreted as a Markov property. Indeed, consider mutations which
occur continuously instead of a single jump and the motion down the ranks is
terminated at a random moment. The Markov property requires that the
termination moment depends only on the current location (and not on the
state of origin of the mutation). With this property, equation (\ref{Gibbs3}%
) become almost obvious: in order to reach the state $y$ from $y^{\circ }$,
a mutation needs to reach first the intermediate state $y^{\prime }$ and
then proceed further to $y$. Differentiation of (\ref{Gibbs3}) with respect
to $y$ yields 
\begin{equation}
f_{\zeta }(y,y^{\prime })=\frac{f_{\zeta }(y,y^{\circ })}{F_{\zeta
}(y^{\prime },y^{\circ })}  \label{Gibbs3a}
\end{equation}

Another useful interpretation is treating each mutation as a sequence $%
y_{0},y_{1},...,y_{k}$ of a large number $k$ of small and stochastically
independent steps $\Delta y_{i}=y_{i-1}-y_{i},$ $i=1,...,k$ so that $%
y_{0}=y^{\circ }$ is the mutation origin and $y_{k}=y$ is its final state.
Each step $\Delta y_{i}$ is associated with effective energy $u_{i}$ so that
step $i$ occurs with probability $\lambda _{i}=\exp (-u_{i})$ when and if
the previous step $i-1$ is completed. If the step $i-1$ does not occur, then 
$u_{i}$ is taken to be infinite and the probability of step $i$ is zero.
Hence the probability of reaching $y_{k}$ and beyond is given by the
exponent 
\begin{equation}
F_{\zeta }(y_{k},y_{0})=\prod_{i=1}^{k}F_{\zeta }(y_{i},y_{i-1})=\exp
(-\sum_{i=1}^{k}u_{i})  \label{Gmes}
\end{equation}
which can be interpreted as a Gibbs measure \cite{GibbsMeasure80}. This
demonstrates the existence of a link between Gibbs mutations and Gibbs
measures. Note the functional consistency of the equations (\ref{mut-f0}), (%
\ref{Gibbs3}) and (\ref{Gibbs3a}) and (\ref{Gmes}). As Gibbs measures, Gibbs
mutations can be introduced on mathematical trees but we can approach more
complex cases with a more general (but perhaps less physically transparent)
definition.

Preferential mixing and mutations in more complex geometries need a more
general approach to Gibbs mutations. The general definition of Gibbs
mutations is given by the following expression for $f_{\zeta }(\mathbf{y},%
\mathbf{y}^{\circ })$ 
\begin{equation}
f_{\zeta }(\mathbf{y},\mathbf{y}^{\circ })=\frac{f_{0}(\mathbf{y})\Psi (%
\mathbf{y},\mathbf{y}^{\circ })}{h_{0}^{+}(\mathbf{y}^{\circ })}H(\mathbf{y}%
^{\circ },\mathbf{y})  \label{mut-tc}
\end{equation}
where $h_{0}^{+}(\mathbf{y}^{\circ })=h^{+}(\mathbf{y}^{\circ },[f_{0}])$ is
introduced. These mutations are non-positive ($\mathbf{y}+\mathbf{\zeta \;}%
\preccurlyeq \mathbf{y}$). Equation (\ref{mut-f0}) can be easily recovered
from (\ref{mut-tc}) under simplifying assumptions. The H-theorem given in
Appendix \ref{sA5} demonstrates that the function $f_{0}$ in (\ref{mut-f0})
and in (\ref{mut-tc}) is, in fact, the equilibrium distribution specified by
(\ref{eq-f0}).

\subsection{Fluctuation theorem}

Consider an isolated thermodynamic system, which is characterised by the
entropy $\hat{s}(y)$ and by corresponding equilibrium distribution 
\begin{equation}
\hat{f}(y)\sim \exp \left( \hat{s}(y)\right)
\end{equation}
The system changes its states though a random walk between the states $%
...,y_{-1},y_{0},y_{1},...$ by moving from the current node $i$ to its
neighbour $i-1$ or $i+1$ with probabilities $\ \alpha _{i}^{-}$ and $\alpha
_{i}^{+}$ correspondingly. The detailed balance at equilibrium requires that 
\begin{equation}
\hat{f}_{i}\alpha _{i}^{+}=\hat{f}_{i+1}\alpha _{i-1}^{-},\;\;\hat{f}%
_{i}\sim \exp \left( \hat{s}_{i}\right)  \label{AD_eq}
\end{equation}
where $\hat{f}_{i}$ is the discrete version of the equilibrium distribution
and $\hat{s}_{i}=\hat{s}(y_{i})$. The condition (\ref{AD_eq}) then takes the
form 
\begin{equation}
\frac{\alpha _{i}^{+}}{\alpha _{i+1}^{-}}=\frac{\hat{f}_{i+1}}{\hat{f}_{i}}%
=\exp \left( \Delta \hat{s}_{i}\right) ,\;\;\Delta \hat{s}_{i}=\hat{s}_{i+1}-%
\hat{s}_{i}
\end{equation}
Let $\Bbb{P}(j^{\circ }\rightarrow j,k)$ be the probability of transition
from node $j^{\circ }$ to node $j$ after exactly $k$ steps (assuming of
course that $\left| j-j^{\circ }\right| \leq k$). Each given sequence $%
j^{\circ }\rightarrow ...\rightarrow j$ can be reversed $j\rightarrow
...\rightarrow j^{\circ }$ and we may assume $j\geq j^{\circ }$ without loss
of generality. The ratio of probabilities of these direct and reverse
sequences is now evaluated. The direct sequence has $j-j^{\circ }$
increasing steps ($i\rightarrow i+1$) and, possibly, a certain number of
closed loops (i.e. subsequences that start and finish at the same location)
while the reverse sequence has the $j-j^{\circ }$ of similar but decreasing
steps ($i+1\rightarrow i$) and the exactly same number of the exactly same
close loops as in the direct sequence. The ratio of probabilities becomes 
\[
\frac{\Bbb{P}\left( j^{\circ }\rightarrow j,k\right) }{\Bbb{P}\left(
j\rightarrow j^{\circ },k\right) }=\frac{\Bbb{P}_{\Sigma 0}}{\Bbb{P}_{\Sigma
0}}\frac{\Bbb{P}\left( j^{\circ }\rightarrow j,j-j^{\circ }\right) }{\Bbb{P}%
\left( j\rightarrow j^{\circ },j-j^{\circ }\right) }= 
\]
\begin{equation}
=\prod_{i=j^{\circ }}^{j-1}\frac{\alpha _{i}^{+}}{\alpha _{i+1}^{-}}=\exp
\left( \sum_{i=j^{\circ }}^{j-1}\Delta \hat{s}_{i}\right) =\exp \left( \hat{s%
}_{j}-\hat{s}_{j^{\circ }}\right)
\end{equation}
where $\Bbb{P}_{\Sigma 0}$ represents the probability of the all closed
loops occurring in the sequence. Note that the ratio does not depend on the
path and on $k.$ The ratio then can be written in terms of probabilities of
changing from entropy $\hat{s}^{\circ }$ to entropy $\hat{s}$ and back 
\begin{equation}
\frac{\Bbb{P}\left( \hat{s}^{\circ }\rightarrow \hat{s}\right) }{\Bbb{P}%
\left( \hat{s}\rightarrow \hat{s}^{\circ }\right) }=\exp \left( \hat{s}-\hat{%
s}^{\circ }\right)  \label{AD_FT}
\end{equation}
This equation is essentially the statement of the fluctuation theorem \cite
{Ftheor2002} applied to the case under consideration.

The fluctuation theorem evaluates only the ratio of probabilities but not
the probabilities $\Bbb{P}\left( \hat{s}^{\circ }\rightarrow \hat{s}\right) $
and $\Bbb{P}\left( \hat{s}\rightarrow \hat{s}^{\circ }\right) $ themselves.
It is applicable to any stochastic process that can be considered as a limit
of the random walk specified here. In case of a simple random walk the
probabilities $\Bbb{P}\left( \hat{s}^{\circ }\rightarrow \hat{s}\right) $
become normal after a large number of steps, in accordance with the central
limit theorem. This, however, does not have to be the case for more general
walks as considered below.

\subsection{Near-Gibbs mutations with exponential distributions}

The mutations analysed in this subsection are the near-Gibbs mutations,
which are similar to Gibbs mutations but may be infrequently positive, that
is $\acute{y}=y+\zeta \preccurlyeq y$ in most cases but $\acute{y}\succ y$
is also occasionally possible. Consider mutations that are generated by a
random walk with a termination after a randomly chosen number of steps. We
demonstrate that under some conditions discussed below, the probability
distribution of these mutations becomes exponential and specified by (\ref
{D-exp}).

Let the mutations be represented by a random walk with stochastically
similar steps and termination probability given by $1-\lambda ,$ where $%
0<\lambda <1.$ The probability of each consequent time step is then $\lambda
=\exp (-u),$ where the exponential form with $u$ is used according to the
notations accepted for the Gibbs measures (\ref{Gmes}). The mutation $\zeta $
can be represented by the sums 
\begin{equation}
\zeta =\sum_{i=0}^{m}\xi _{i}
\end{equation}
where all $\xi _{i}$ are stochastically equivalent and independent and $m$
is a random integer value with the probability distribution 
\begin{equation}
\Bbb{P}\left( m=k\right) =\lambda ^{k}(1-\lambda )
\end{equation}
The variable $\zeta $ represents mutations $y=y^{\circ }+\zeta $ originated
at $y^{\circ }$.

The moment producing function of the random steps is given by 
\begin{equation}
\Upsilon _{\xi }(b)=\left\langle \exp \left( b\xi \right) \right\rangle
=\sum_{i=0}^{\infty }\frac{\left\langle \xi ^{i}\right\rangle }{i!}b^{i}
\end{equation}
The moment producing function of $k$ independent steps is 
\begin{equation}
\Upsilon _{k}(b)=\left\langle \exp \left( b\sum_{i=0}^{k}\xi _{i}\right)
\right\rangle =\left\langle \exp \left( b\xi \right) \right\rangle
^{k}=\left( \Upsilon _{\xi }(b)\right) ^{k}
\end{equation}
The moment producing function for the variable $\zeta $ representing
mutations is given by 
\[
\frac{\Upsilon _{\zeta }(b)}{1-\lambda }=\frac{\left\langle \exp \left(
b\zeta \right) \right\rangle }{1-\lambda }=1+\lambda \Upsilon
_{1}(b)+\lambda ^{2}\Upsilon _{2}(b)+...= 
\]
\begin{equation}
=\sum_{i=0}^{\infty }\left( \lambda \Upsilon _{\xi }(b)\right) ^{i}=\frac{1}{%
1-\lambda \Upsilon _{\xi }(b)}
\end{equation}
The value of $\lambda $ is presumed to be closed to $1$ that is $u$ is
small. In this case the sequence involves many steps before termination and
the universal limiting distribution of $\zeta $ can be determined. The
function $\Upsilon _{\xi }(b)$ can be expanded into a series to give 
\begin{equation}
\Upsilon _{\zeta }(b)=\frac{1/\lambda -1}{1/\lambda -1-\left\langle \xi
\right\rangle b-\frac{\left\langle \xi ^{2}\right\rangle }{2}b^{2}-...}
\label{AD_exp}
\end{equation}
With the use of 
\begin{equation}
1/\lambda -1=\exp (u)-1=u+...
\end{equation}
we observe that characteristic values of $b$ in (\ref{AD_exp}) are small
when $u$ is small. The leading order expression for the moment generating
function becomes 
\begin{equation}
\Upsilon (b)=\frac{1}{1-\frac{\left\langle \xi \right\rangle }{u}b}
\end{equation}
which means that 
\begin{equation}
f(\zeta )=b_{1}\exp \left( b_{1}\zeta \right) ,\;\;b_{1}=-\frac{u}{%
\left\langle \xi \right\rangle }
\end{equation}
Without loss of generality, we imply that $\left\langle \xi \right\rangle $
is negative and $\zeta $ is predominantly negative while $b_{1}>0$.

In special cases, when $\left| \left\langle \xi \right\rangle \right| $ is
zero or small, equation for $b_{1}$ needs to be replaced by another equation 
\begin{equation}
b_{1}=\frac{-\left| \left\langle \xi \right\rangle \right| +\sqrt{%
\left\langle \xi \right\rangle ^{2}+2u\left\langle \xi ^{2}\right\rangle }}{%
\left\langle \xi ^{2}\right\rangle }
\end{equation}
which is obtained by retaining the second moment $\left\langle \xi
^{2}\right\rangle $ in (\ref{AD_exp}). If $\left\langle \xi \right\rangle =0$
then $b_{1}=(2u/\left\langle \xi ^{2}\right\rangle )^{1/2}$ and the exponent
for the positive mutations must be the same $b_{2}=b_{1}$ since the problem
with $\left\langle \xi \right\rangle =0$ is symmetric with respect to the
substitution of $-\zeta $ for $\zeta $.

If $\left\langle \xi \right\rangle $ is not small then $b_{2}$ becomes much
larger than $b_{1}$ and evaluation of the distribution on the positive side
becomes difficult --- at the leading order the positive fluctuations are too
small to be detected. The fluctuation theorem, nevertheless, allows to
determine the probability distribution of positive mutations even if these
mutations are very small and infrequent. Indeed, we may write the
fluctuation theorem (\ref{AD_FT}) in the form 
\begin{equation}
\frac{\Bbb{P}\left( \zeta =\zeta ^{\circ }\right) }{\Bbb{P}\left( \zeta
=-\zeta ^{\circ }\right) }=\exp \left( -\hat{\beta}\zeta ^{\circ }\right)
,\;\;-\frac{d\hat{s}}{dy}=\hat{\beta}  \label{AD_FT2}
\end{equation}
resulting in 
\begin{equation}
f_{\zeta }(\zeta )=b_{0}\exp \left( -b_{2}\zeta \right) ,\;\zeta \geq 0
\end{equation}
\begin{equation}
b_{2}=b_{1}+\hat{\beta}
\end{equation}
as suggested in equation (\ref{D-exp}). The value $\hat{\beta}$, which is
expected to be positive for predominantly negative mutations, needs to be
connected to the stochastic properties of the variable $\xi $. We can apply
the fluctuation theorem to the distribution of $\xi $%
\begin{equation}
\frac{\Bbb{P}\left( \xi =\xi ^{\circ }\right) }{\Bbb{P}\left( \xi =-\xi
^{\circ }\right) }=\exp \left( -\hat{\beta}\xi ^{\circ }\right)
\end{equation}
which after multiplying by $\Bbb{P}\left( \xi =-\xi ^{\circ }\right) $ and
integrating over all $\xi ^{\circ }$ yields \cite{Ftheor2002} 
\begin{equation}
\left\langle \exp \left( \hat{\beta}\xi \right) \right\rangle =1
\label{AD_FTexp}
\end{equation}
This equation allows to determine $\hat{\beta}$. If $\xi $ is small, then
expanding the exponent in (\ref{AD_FTexp}) into a power series allows us to
obtain the following explicit expression for $\hat{\beta}:$%
\begin{equation}
\hat{\beta}=-2\frac{\left\langle \xi \right\rangle }{\left\langle \xi
^{2}\right\rangle }
\end{equation}
Note that $\hat{\beta}$ is positive when $\xi $ is predominantly negative.

\begin{figure}[ht] 
\caption{Simulated (solid line) and predicted (symbols) double exponential distribution of mutations 
obtained as a limit of the random walk with interruptions. 
The bottom figure shows the same distribution as the top figure but using a logarithmic ordinate.} 
\begin{center} 
\includegraphics[width=10cm]{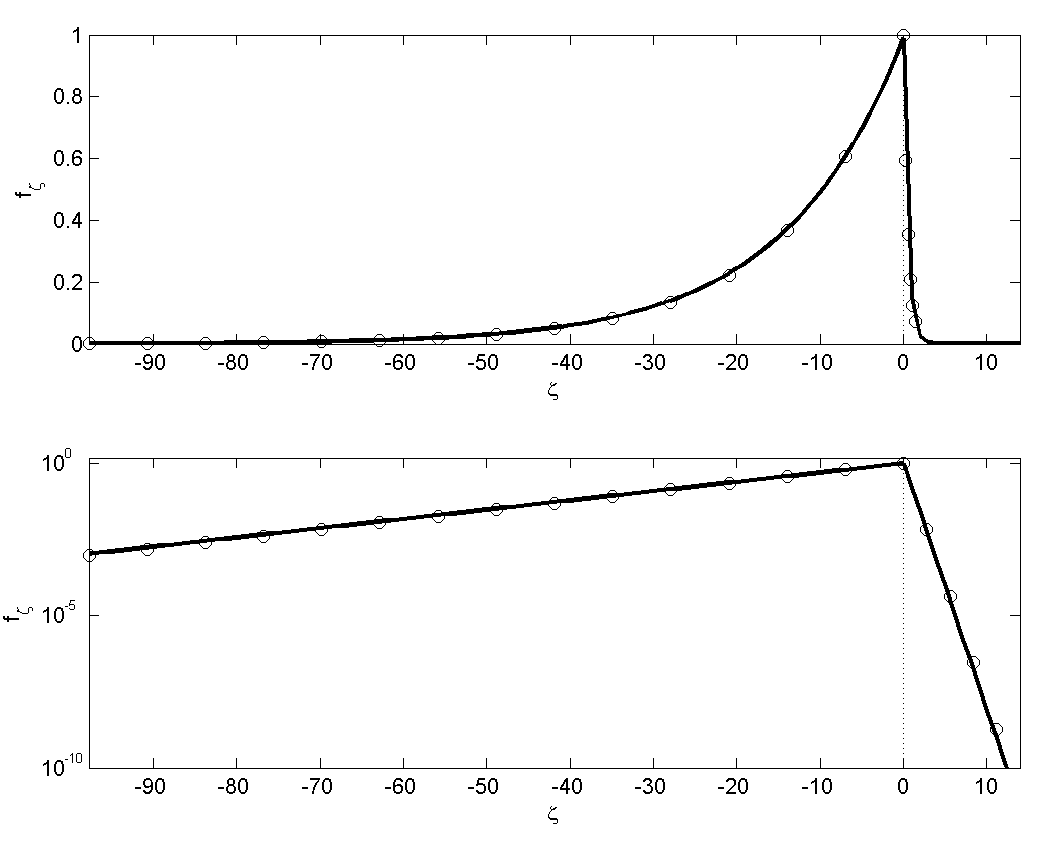}  
\label{fig_Ap1}  
\end{center}
\end{figure}

Convergence of a random walk distribution to the double-exponential
distribution is shown in Figure \ref{fig_Ap1}. The walk involves random steps%
$\ \xi =-1$ with probability 85\% and $\xi =+1$ with probability 15\%, while
the termination parameter was set to $\lambda =0.95$. The solid line
presents the numerical evaluation while the symbols show the expected
double-exponential distribution. This distribution has around 4\% positive
mutations.

\section{Evolution of competitive systems\label{sA5}}

\subsection{Abstract competition \label{sA4}}

Consider a large number of autonomous elements that possess two sets of
properties a) non-conservative (information-like), b) conservative
(energy-like) and, possibly, c) a set of physical coordinates. Abstract
competition involves the following steps:

\begin{enumerate}
\item  \textbf{Selection.} Random selection of elements to form competing
couples, possibly with some preferences and/or isolations and possibly from
the same locality in physical space.

\item  \textbf{Competition.} Determining the winner and the loser for each
couple on the basis of the element properties, possibly with some randomness.

\item  \textbf{Conservative mixing. }The conservative properties are
redistributed on even basis or from the loser to the winner. The total
amounts are preserved in this redistribution.

\item  \textbf{Non-conservative mixing.} The non-conservative properties are
redistributed from the winner to the loser so that some or all
non-conservative properties of the loser are lost.

\item  \textbf{Mutations.} The redistribution of non-conservative properties
may involve random changes (i.e. mutations), which are expected to be mostly
detrimental to competitiveness of the elements.
\end{enumerate}

It appears that the abstract competition can be naturally represented by a
system of Pope particles with competitive and conservative mixing. In the
present work, the conservative properties are limited to the particles
themselves, i.e. each particle has the conservative property of $1.$
Particles with finite life times can be considered if needed.

\subsection{Competitive pdf equation}

In the presence of mutations, the evolution equation for the pdf takes the
form 
\begin{equation}
\frac{\partial f(\mathbf{y)}}{\partial t}=\gamma \left( \Bbb{M}\left( h^{+}(%
\mathbf{y},[f])f(\mathbf{y})\right) -h^{-}(\mathbf{y},[f])f(\mathbf{y}%
)\right)   \label{eqm}
\end{equation}
where and $\Bbb{M}$ is the mutation operator 
\begin{equation}
\Bbb{M}\left( f(\mathbf{y})\right) =\stackunder{\infty }{\int }f_{\zeta }(%
\mathbf{y,y^{\circ })}f(\mathbf{y^{\circ })}d\mathbf{y}^{\circ },
\end{equation}
$f_{\zeta }(\mathbf{y,y^{\circ })}$ is the normalised distribution of
mutations originated from state $\mathbf{y}^{\circ }$ and $\gamma $ is the
constant determining the overall rate of mixing. The non-homogeneous terms
present in the standard pdf equation (\ref{pdfx}) are not included here
since the case is presumed to be spatially homogeneous. This equation is a
generalisation of the evolution equation derived in Ref. \cite{K-PS2010} and
is quite transparent: $h^{+}(\mathbf{y},[f])f(\mathbf{y})$ is the total
fraction of losers to the element $\mathbf{y}$ and $h^{-}(\mathbf{y},[f])f(%
\mathbf{y})$ is the total fraction of winners over the element $\mathbf{y}$.
The losers are subject to mutations originated at $\mathbf{y}$ while the
winners remove particles from location $\mathbf{y}$.

\subsection{Mutation-free evolution}

In the absence of mutations the rate of change of the distribution can be
easily expressed in terms of relative ranking 
\begin{equation}
\frac{\partial \ln (f(\mathbf{y))}}{\partial t}=\gamma \theta (\mathbf{y,}%
[f]),\;\;\;\mathbf{\zeta }=0  \label{c-e}
\end{equation}
where $\gamma $ is the constant determining the rate of mixing and $\theta (%
\mathbf{y,}[f])=r(\mathbf{y,}[f])\psi (\mathbf{y,}[f])$. We note that the
cases where there is a draw do not change the properties when mutations are
not present.

In case of transitive competition of non-isolated particles, a steady state
is possible only in a trivial case when all particles have the same rank
(and thus $R_{pq}=0$ for any $p$ and $q$). Indeed, any particle $p$, which
has a rank $r_{p}$ lower than that of the leader $r_{\ast }$, would
eventually lose competition to the leader and acquire the leading rank $%
r_{\ast }$. The steady state is reached only when all particles are ranked
at $r_{\ast }$. A steady state with $R_{pq}\neq 0$ is, however, possible for
intransitive competition. For example, consider particles equally
distributed between the three states in the scissors-paper-rock competition.
This distribution is steady as the probabilities of losing a particle and
gaining a particle are the same for all three states. Stability of
non-trivial steady distributions in intransitive competitions is not
guaranteed and needs to be examined.

Equation (\ref{c-e}) indicates that in absence of mutations any stationary
solution requires $\theta (\mathbf{y,}[f_{0}])=0$ for any $\mathbf{y}$ such
that $f(\mathbf{y})>0$. Consider variation $\delta f(\mathbf{y})=f(\mathbf{y}%
)-f_{0}(\mathbf{y})$, whose support is not larger than that of $f_{0}(%
\mathbf{y})$ (that is $\delta f(\mathbf{y})=0$ \ for all $\mathbf{y}$ that$\
f(\mathbf{y})=0$) then within the support of $f_{0}(\mathbf{y})$ we have $%
\theta (\mathbf{y},[f_{0}])=0$ and 
\begin{equation}
\theta (\mathbf{y},[f])=\theta (\mathbf{y},[f_{0}+\delta f])=\theta (\mathbf{%
y},[f_{0}])+\theta (\mathbf{y},[\delta f])=\theta (\mathbf{y},[\delta f])
\end{equation}
The linearised form of equation (\ref{c-e}) is 
\begin{equation}
\frac{\partial \delta f(\mathbf{y})}{\partial t}=\gamma f_{0}(\mathbf{y}%
)\theta (\mathbf{y,}[\delta f])
\end{equation}
With the use of $\widetilde{\delta f}(\mathbf{y})=\delta f(\mathbf{y})/f_{0}(%
\mathbf{y})^{1/2}$ this equation can be written as 
\begin{equation}
\frac{\partial \widetilde{\delta f}}{\partial t}=\gamma \Bbb{L}_{f}%
\widetilde{\delta f}  \label{ddfQ}
\end{equation}
where $\Bbb{L}_{f}$ is a linear operator applied to $\widetilde{\delta f}$
and defined by 
\begin{equation}
\Bbb{L}_{f}\widetilde{\delta f}(\mathbf{y})=\stackunder{\infty }{\int }\Psi (%
\mathbf{y},\mathbf{y}^{\prime })R(\mathbf{y},\mathbf{y}^{\prime })\widetilde{%
\delta f}(\mathbf{y}^{\prime })\sqrt{f(\mathbf{y})f(\mathbf{y}^{\prime })}d%
\mathbf{y}^{\prime }
\end{equation}
The positiveness of $f_{0}(\mathbf{y})$ for all $\mathbf{y}$ where $\delta f(%
\mathbf{y})\neq 0$ must be noted. The operator $\Bbb{L}_{f}$ is
skew-symmetric (anti-symmetric) 
\begin{equation}
\stackunder{\infty }{\int }\widetilde{\delta f}(\mathbf{y})\Bbb{L}_{f}%
\widetilde{\delta f}(\mathbf{y})d\mathbf{y}=0
\end{equation}
due to $R(\mathbf{y},\mathbf{y}^{\prime })=-R(\mathbf{y}^{\prime },\mathbf{y}%
)$. The solution of equation (\ref{ddfQ}) with initial conditions \ $%
\widetilde{\delta f}(\mathbf{y})=\widetilde{\delta f^{\circ }}(\mathbf{y})$
at $t=0$ is given by 
\begin{equation}
\widetilde{\delta f}(\mathbf{y})=\exp (\gamma t\Bbb{L}_{f}\Bbb{)}\widetilde{%
\delta f^{\circ }}(\mathbf{y})
\end{equation}
where the exponential operator is unitary. Hence, unless $R_{pq}=0$ for all
particle couples $p$ and $q,$ the system of competing particles without
mutations has non-decaying oscillations. Generally, negative mutations tend
to act as a stabilising factor while positive mutations tend to be
destabilising.

\subsection{Nearly mutation-free evolution}

Consider the overall steady distribution $f(\mathbf{y})$, represented by a
superposition of $a_{I}\phi _{I}(\mathbf{y}),$ where $\phi _{I}(\mathbf{y})$
is the distribution of particles within one of the subsystems $I=1,...,K$
located in domains $\frak{D}_{I}$ as specified by (\ref{grp_f}). The major
changes in the system occur due to the competition between the subsystems,
which does not involve mutations, while smaller adjustments of distributions
happen due to competitions within each subsystem and may involve some
non-positive mutations. The distribution $f(\mathbf{y})$ is, initially, at
equilibrium.

If competition is transitive and the subsystem leaders $\mathbf{y}_{I\ast }$
are not isolated from each other, then any non-trivial steady distribution
is impossible. Indeed, once a subsystem leader has lost competition to a
leader from another subsystem, it cannot be replaced due to absence of
mutations between subsystems and absence of positive mutations within the
subsystem. The system keeps evolving until only one subsystem is left. A
non-trivial steady state, however, is possible in intransitive competition.

Let us consider linear stability of the equilibrium with respect to the
disturbance, which is represented by small changes in $a_{I}$ while $\phi
_{I}(\mathbf{y})$ are not changed initially and remain the same within
accuracy of our analysis. In the absence of mutational exchanges between the
subsystem, this leads to the relation 
\begin{equation}
\frac{\partial a_{I}}{\partial t}=\gamma \bar{\Theta}_{I}([f])a_{I}=\gamma
\sum_{J}a_{I}a_{J}\bar{\Theta}_{IJ}  \label{eqg}
\end{equation}
which can be obtained from (\ref{eqm}) by integrating $f$ over each of the
subsystem domains $\frak{D}_{I}$ under assumptions that $\phi _{I}(\mathbf{y}%
)$ (and consequently $\bar{\Theta}_{IJ})$ are at equilibrium and do not
change in time and that domains $\frak{D}_{I}$ do not exchange mutations.

Since $f(\mathbf{y})$ is steady, equation (\ref{eqg}) indicates that $\bar{%
\Theta}_{I}a_{I}=0$. We investigate stability of the system with respect to
changes of every $a_{I}\neq 0$ and $\bar{\Theta}_{I}=0$ by $\Delta a_{I}$
assuming that $\phi _{I}(\mathbf{y})$ do not change that is $\Delta f(%
\mathbf{y})=\Sigma _{I}\phi _{I}(\mathbf{y})\Delta a_{I}.$ Note that if $%
a_{J}=0$, competition without mutations does not allow to change this value.
Equation (\ref{eqg}) takes the form 
\begin{equation}
\frac{\partial a_{I}}{\partial t}=\gamma a_{I}\Delta \bar{\Theta}_{I}=\gamma
a_{I}\sum_{J=1}^{K}\bar{\Theta}_{IJ}\Delta a_{J}
\end{equation}
With $\widetilde{\Delta a}_{I}=\Delta a_{I}/a_{I}^{1/2}$ the linearised
version of this equation takes the form 
\begin{equation}
\frac{\partial \widetilde{\Delta a}_{I}}{\partial t}=\sum_{J=1}^{K}L_{IJ}%
\widetilde{\Delta a}_{J}=\Bbb{L}_{a}\widetilde{\Delta a}_{I}  \label{ddaQ}
\end{equation}
where the linear operator $\Bbb{L}_{a}$ is represented by the skew-symmetric
matrix 
\begin{equation}
L_{IJ}=a_{I}^{1/2}\bar{\Theta}_{IJ}a_{J}^{1/2}
\end{equation}
Note that $a_{I}>0$ for any $I$ that $\Delta a_{I}>0$. The solution of
equation (\ref{ddaQ}) with initial conditions $\widetilde{\Delta a}_{I}=%
\widetilde{\Delta a_{I}^{\circ }}$ at $t=0$ is 
\begin{equation}
\widetilde{\Delta a}_{I}=\sum_{J=1}^{K}\exp (\gamma t\Bbb{L}_{a})\widetilde{%
\Delta a_{I}^{\circ }}
\end{equation}
where the matrix exponent $\exp (\gamma t\Bbb{L}_{a})$ produces a unitary
matrix. Hence, unless $\bar{\Theta}_{IJ}=\bar{R}_{IJ}\bar{\Psi}_{IJ}=0$ for
every $I$ and $J,$ a deviation from the steady state is expected to generate
oscillations in the system. Equilibration of two non-isolated subsystems $I$
and $J$ in the absence of mutational exchanges between the subsystems needs $%
\bar{R}_{IJ}=0$ rather than equivalence of the competitive potentials.


\subsection{Competitive H-theorem}

The competition considered here can be transitive or intransitive,
preferential or non-preferential. Mutations are represented by Gibbs
mutations satisfying formula (\ref{mut-tc}).

\begin{theorem}
\label{thH1}The total entropy of a competitive system with non-positive
Gibbs mutations monotonically increases during evolution of the system until
it reaches its maximal value at the equilibrium.
\end{theorem}

For mutations distributed according to equation (\ref{mut-tc}), the
governing equation (\ref{eqm}) takes the form 
\[
\frac{\partial f(\mathbf{y}_{1})}{\partial t}= 
\]
\[
\gamma \stackunder{\infty }{\int }H(\mathbf{y}_{2},\mathbf{y}_{1})\Psi (%
\mathbf{y}_{1},\mathbf{y}_{2})\frac{f(\mathbf{y}_{2})}{h_{0}^{+}(\mathbf{y}%
_{2})}\left( f_{0}(\mathbf{y}_{1})h^{+}(\mathbf{y}_{2},[f])-f(\mathbf{y}%
_{1})h_{0}^{+}(\mathbf{y}_{2})\right) d\mathbf{y}_{2} 
\]
\begin{equation}
=\gamma \stackunder{\infty }{\int \int }\Phi _{3}(\mathbf{y}_{1},\mathbf{y}%
_{2},\mathbf{y}_{3})\stackunder{\varphi (\mathbf{y}_{1},\mathbf{y}_{3})}{%
\underbrace{\left( f_{0}(\mathbf{y}_{1})f(\mathbf{y}_{3})-f(\mathbf{y}%
_{1})f_{0}(\mathbf{y}_{3})\right) }}d\mathbf{y}_{3}d\mathbf{y}_{2}
\label{eqhm}
\end{equation}
where 
\[
\Phi _{3}(\mathbf{y}_{1},\mathbf{y}_{2},\mathbf{y}_{3})=H(\mathbf{y}_{2},%
\mathbf{y}_{1})\Psi (\mathbf{y}_{1},\mathbf{y}_{2})H(\mathbf{y}_{2},\mathbf{y%
}_{3})\Psi (\mathbf{y}_{3},\mathbf{y}_{2})\frac{f(\mathbf{y}_{2})}{h_{0}^{+}(%
\mathbf{y}_{2})}\geq 0 
\]
Evaluation of the time derivative for the entropy defined by 
\begin{equation}
S([f(\mathbf{y})])=\;-\int_{\infty }f(\mathbf{y})\ln \left( \frac{f(\mathbf{y%
})}{f_{0}(\mathbf{y})}\right) d\mathbf{y+}C_{0}([f_{0}(\mathbf{y})])
\label{A-entrop}
\end{equation}
results, after substitution of (\ref{eqhm}), in 
\[
\frac{dS}{dt}=-\stackunder{\infty }{\int }\frac{\partial f(\mathbf{y}_{1})}{%
\partial t}\ln \left( \frac{f(\mathbf{y}_{1})}{f_{0}(\mathbf{y}_{1})}\right)
d\mathbf{y}_{1}= 
\]
\[
-\gamma \stackunder{\infty }{\int \int \int }\Phi _{3}(\mathbf{y}_{1},%
\mathbf{y}_{2},\mathbf{y}_{3})\stackunder{\Phi _{2}(\mathbf{y}_{1},\mathbf{y}%
_{3})}{\underbrace{\varphi (\mathbf{y}_{1},\mathbf{y}_{3})\ln \left( \frac{f(%
\mathbf{y}_{1})}{f_{0}(\mathbf{y}_{1})}\right) }}d\mathbf{y}_{1}d\mathbf{y}%
_{2}d\mathbf{y}_{3}= 
\]
\begin{equation}
-\gamma \stackunder{\infty }{\int \int \int }\Phi _{3}(\mathbf{y}_{1},%
\mathbf{y}_{2},\mathbf{y}_{3})\frac{\Phi _{2}(\mathbf{y}_{1},\mathbf{y}%
_{3})+\Phi _{2}(\mathbf{y}_{3},\mathbf{y}_{1})}{2}d\mathbf{y}_{1}d\mathbf{y}%
_{2}d\mathbf{y}_{3}
\end{equation}
Here, we use the symmetric properties $\Phi _{3}(\mathbf{y}_{1},\mathbf{y}%
_{2},\mathbf{y}_{3})=\Phi _{3}(\mathbf{y}_{3},\mathbf{y}_{2},\mathbf{y}_{1})$
of the function $\Phi _{3}$ while noting that the variables $\mathbf{y}_{1}$
and $\mathbf{y}_{3}$ are dummy integration variables and can be swapped in
the integral. With the use the expression 
\[
\Phi _{2}(\mathbf{y}_{1},\mathbf{y}_{3})+\Phi _{2}(\mathbf{y}_{3},\mathbf{y}%
_{1})= 
\]
\[
\varphi (\mathbf{y}_{1},\mathbf{y}_{3})\left( \ln \left( \frac{f(\mathbf{y}%
_{1})}{f_{0}(\mathbf{y}_{1})}\right) -\ln \left( \frac{f(\mathbf{y}_{3})}{%
f_{0}(\mathbf{y}_{3})}\right) \right) = 
\]
\begin{equation}
-f_{0}(\mathbf{y}_{1})f(\mathbf{y}_{3})\left( \omega -1\right) \ln (\omega
)\leq 0
\end{equation}
and antisymmetry of the function $\varphi $ 
\[
\varphi (\mathbf{y}_{1},\mathbf{y}_{3})=-\varphi (\mathbf{y}_{3},\mathbf{y}%
_{1}) 
\]
while introducing $\omega $ by 
\[
\omega =\frac{f(\mathbf{y}_{3})}{f_{0}(\mathbf{y}_{3})}\frac{f_{0}(\mathbf{y}%
_{1})}{f(\mathbf{y}_{1})} 
\]
and noting that $\left( \omega -1\right) \ln (\omega )\geq 0,$ we obtain the
relation 
\[
\frac{\gamma }{2}\stackunder{\infty }{\int \int \int }\Phi _{3}(\mathbf{y}%
_{1},\mathbf{y}_{2},\mathbf{y}_{3})f_{0}(\mathbf{y}_{1})f(\mathbf{y}%
_{3})\left( \omega -1\right) \ln (\omega )d\mathbf{y}_{1}d\mathbf{y}_{2}d%
\mathbf{y}_{3}= 
\]
\begin{equation}
=\frac{dS}{dt}\geq 0
\end{equation}
proving the H-theorem.

\subsection{Convergence theorem for transitive competition}

The mutations considered in this subsection are deemed to be general
non-positive mutations (rather than Gibbs mutations), although our analysis
is restricted to transitive competition. Consider competition within a
closed domain $\frak{D}$ that has a finite measure (i.e. volume) $\mu (\frak{%
D})<\infty $ and assume that the mutation distribution function $f_{\zeta }(%
\mathbf{y},\mathbf{y}^{\circ })$ is continuous in this domain. The mixing is
assumed to be non-preferential ($\Psi =1$). The absolute ranking $r_{\#}(%
\mathbf{y})$ is continuous and satisfies additional continuity requirement $%
\mu (\frak{D}_{2}(r_{1},r_{1}+\Delta r))\rightarrow 0$ as $\Delta
r\rightarrow 0$ uniformly in $\frak{D.}$ Here we define $\ \frak{D}%
_{2}(r_{1},r_{2})=\{\mathbf{y|}r_{1}\mathbf{\leq }r_{\#}(\mathbf{y})\mathbf{%
\leq }r_{2}\}$ and, analogously, $\frak{D}_{2}(\mathbf{y}_{1}\mathbf{,y}%
_{2})=\{\mathbf{y}|\mathbf{y}_{1}\mathbf{\preccurlyeq y\preccurlyeq y}%
_{2}\}. $ For the mathematical space of continuous functions $f(\mathbf{y})$
defined on $\frak{D}$ and conventionally denoted by $C^{0}(\frak{D),}$\ the
norm of $f$ can be specified by 
\begin{equation}
\left\| f\right\| =\max_{\mathbf{y}\in \frak{D}}\left( \left| f(\mathbf{y}%
)\right| \right)
\end{equation}
The mathematical restrictions considered here are introduced to keep the
proof transparent and, in principle, can be modified or relaxed.

\begin{theorem}
\label{thH2}If competition is transitive and mutations are non-positive, the
system asymptotically approaches its equilibrium state where the total
entropy of the system reaches its maximal value.
\end{theorem}

If mutations are non-positive, the governing equation (\ref{eqm}) takes the
form 
\[
\frac{\partial f(\mathbf{y})}{\partial t}= 
\]
\begin{equation}
\gamma \stackunder{\frak{D}}{\int }H(\mathbf{y}^{\circ },\mathbf{y})\left(
f_{\zeta }(\mathbf{y},\mathbf{y}^{\circ })(1-h^{-}(\mathbf{y}^{\circ
},[f]))-f(\mathbf{y})\right) f(\mathbf{y}^{\circ })d\mathbf{y}^{\circ }
\label{prop1}
\end{equation}
implying that for any $\mathbf{y}_{1}$ the evolution of the distribution in
the region $\mathbf{y}\succcurlyeq \mathbf{y}_{1}$ does not depend on the
distribution in the region $\mathbf{y\prec y}_{1}$. If a stationary solution
is reached for $\mathbf{y}\succcurlyeq \mathbf{y}_{1},$ it will remain in
this state although $f(\mathbf{y})$ may still evolve at $\mathbf{y\prec y}%
_{1}$. we can formally rewrite equation (\ref{prop1}) in the form 
\[
\frac{\partial f(\mathbf{y)}}{\partial t}=\int_{\mathbf{y}}^{\mathbf{y}%
_{\ast }}f_{\zeta }(\mathbf{y,y}^{\circ }\mathbf{)}\left( 1-h^{-}(\mathbf{y}%
^{\circ },[f])\right) f(\mathbf{y}^{\circ })d\mathbf{y}^{\circ } 
\]
\begin{equation}
-h^{-}(\mathbf{y},[f])f(\mathbf{y})  \label{prop2}
\end{equation}
where 
\begin{equation}
h^{-}(\mathbf{y,[}f])=\int_{\mathbf{y}}^{\mathbf{y}_{\ast }}f(\mathbf{y}%
^{\circ })d\mathbf{y}^{\circ }
\end{equation}
The limits in this equation imply that the integral is evaluated over the
region $\frak{D}_{2}(\mathbf{y,y}_{\ast })$ where $\mathbf{y}_{\ast }$ is
the location of the leading particle. Note that the regions $\frak{D}_{2}(%
\mathbf{y}_{1}\mathbf{,y}_{1})$ have a zero measure for any $\mathbf{y}_{1}$.

If the steady solution is established in the region $\frak{D}_{2}(\mathbf{y}%
_{1}\mathbf{,y}_{\ast })$ but not in the region $\frak{D}_{2}(\mathbf{y,y}%
_{1}),$ which is presumed to be small as $\mathbf{y}$\ is taken close to $%
\mathbf{y}_{1},$ a steady state must be established in the second region as
it is effectively controlled by the first region and mutations in a small
region can always be deemed close to being Gibbs mutations.

The integral in (\ref{prop2}) can be divided into several terms 
\begin{equation}
\frac{\partial f}{\partial t}=\gamma Q_{0}+\gamma \Bbb{Q}_{1}(f)-\gamma \Bbb{%
Q}_{2}(f)-\gamma h_{0}^{-}(\mathbf{y}_{1})f(\mathbf{y})
\end{equation}
where 
\[
Q_{0}=\int_{\mathbf{y}_{1}}^{\mathbf{y}_{\ast }}f_{\zeta }(\mathbf{y,y}%
^{\circ }\mathbf{)(}1\mathbf{-}h_{0}^{-}(\mathbf{y}^{\circ }))f_{0}(\mathbf{y%
}^{\circ })d\mathbf{y}^{\circ }
\]
\[
\Bbb{Q}_{1}(f)=\int_{\mathbf{y}}^{\mathbf{y}_{1}}f_{\zeta }(\mathbf{y,y}%
^{\circ }\mathbf{)}(1-h^{-}(\mathbf{y}^{\circ },[f]))f(\mathbf{y}^{\circ })d%
\mathbf{y}^{\circ }
\]
\[
\Bbb{Q}_{2}(f)=f(\mathbf{y})\left( h^{-}(\mathbf{y},[f])-h_{0}^{-}(\mathbf{y}%
_{1})\right) =f(\mathbf{y})\int_{\mathbf{y}}^{\mathbf{y}_{1}}f(\mathbf{y}%
^{\circ })d\mathbf{y}^{\circ }
\]
and seek a solution in form of the time steps 
\[
f(\mathbf{y},t+\Delta t)=
\]
\begin{equation}
f(\mathbf{y},t)\left( 1-h_{0}^{-}(\mathbf{y}_{1})\gamma \Delta t\right)
+\gamma \Delta t\left( Q_{0}+\Bbb{Q}_{1}(f)-\Bbb{Q}_{2}(f)\right) 
\label{prop3}
\end{equation}
The norms of the operators $\Bbb{Q}_{1}$ and $\Bbb{Q}_{2}$ can be easily
estimated by 
\[
\left\| \Bbb{Q}_{1}\right\| \leq \left\| f_{\zeta }\right\| _{\Delta }\mu
_{\Delta }
\]
\[
\left\| \Bbb{Q}_{2}\right\| \leq 2\left\| f\right\| _{\Delta }\mu _{\Delta }
\]
where 
\[
\mu _{\Delta }=\mu (\frak{D}_{2}(\mathbf{y,y}_{1}))
\]
is the measure of the region $\frak{D}_{2}(\mathbf{y,y}_{1})$ and the norms
subscribed by $\Delta $ are evaluated over $\frak{D}_{2}(\mathbf{y,y}_{1})$.
The norm of operators $\Bbb{Q}_{1}$ and $\Bbb{Q}_{2}$ can be made smaller
than any given $\varepsilon >0$ by selecting $r_{\#}(\mathbf{y)}$
sufficiently close to (but still at a finite distance from) $r_{\#}(\mathbf{y%
}_{1}).$ \ The leading order estimate of the stationary distribution on $%
\frak{D}_{2}(\mathbf{y,y}_{1})$\ is given by $f_{0}\approx Q_{0}/h_{0}^{-}(%
\mathbf{y}_{1}),$ constraining the norms of the converging distributions
(assuming reasonable initial conditions). Since $Q_{0}$ does not depend on $f
$ and the norms of the operators $\Bbb{Q}_{1}$ and $\Bbb{Q}_{2}$ are
relatively small, the operator on the right-hand side of equation (D.25)
becomes a contraction mapping (provided $\varepsilon $ is selected so that $%
\varepsilon <h_{0}^{-}(\mathbf{y}_{1})$ and the time step $\Delta t$ is
sufficiently small). This proves convergence and uniqueness of the solution.

Note that $\varepsilon $ can be selected uniformly over the domain with the
exception of the vicinity of the leading particle where $h_{0}^{-}(\mathbf{y}%
)\rightarrow 0$ as $\mathbf{y\rightarrow y}_{\ast }$. \ One may also notice
that $\partial f(\mathbf{y)/}\partial t\rightarrow 0$\ as $\mathbf{%
y\rightarrow y}_{\ast }$. The vicinity of leading particle needs special
consideration that takes into account the discrete nature of the particle
system. The position of the leading particle is fixed and we need to
consider the distribution behind the leading particle where $h_{0}^{-}(%
\mathbf{y})\geq 1/n$ is positive, where $n$ is the total number of
particles. The singularity of $h_{0}^{-}(\mathbf{y})\rightarrow 0$ as $%
\mathbf{y\rightarrow y}_{\ast }$ indicates that convergence to a steady
solution in the leading group is not uniform and can be quite slow. Indeed,
let two particles $p$ and $q$ be initially located at $\mathbf{y}_{\ast }.$
This distribution is not at equilibrium since eventually, after certain
characteristic time $t_{\ast },$ these two particles would form a mixing
couple and one of them is destined to leave the leading position. If the
total number of particles $n$ is very large, the deviation from the
equilibrium is small but the time $t_{\ast }$ can be very large.

As $f\rightarrow f_{0}$, the entropy defined by (\ref{A-entrop}) approaches
its maximal value; the theorem, however, does not guarantee that this
process is monotonic.

\end{appendices}



\bigskip

\bigskip

\bigskip

\bigskip

\bigskip

\bigskip

\bigskip

\bigskip


\bigskip


\bigskip

\end{document}